\documentclass[twocolumn,showpacs,superscriptaddress,preprintnumbers,prd]{revtex4-1}
\usepackage{graphicx,amsmath,amssymb,dcolumn,bm}
\usepackage{fixmath}
\usepackage{feynmp}
\usepackage{epic,eepic}
\usepackage{slashed}
\newcommand{\pslash}{\slashed{p}}
\allowdisplaybreaks[2]
\usepackage{calrsfs}
\DeclareMathAlphabet{\pazocal}{OMS}{zplm}{m}{n}

\begin{document}
\preprint{KEK-TH-2575}
\title{Pion-production cross sections in neutrino reactions \\
for studying generalized parton distributions of the nucleon}
\author{Xurong Chen}
\affiliation
{Institute of Modern Physics, Chinese Academy of Sciences, Lanzhou 730000, China}
\affiliation
{University of Chinese Academy of Sciences, Beijing 100049, China}
\affiliation
{South China Normal University, Guangzhou 510006, China}
\author{S. Kumano}
\affiliation
  {Department of Mathematics, Physics, and Computer Science,
   Faculty of Science, Japan Women's University,\\           
   Mejirodai 2-8-1, Bunkyo-ku, Tokyo 112-8681, Japan}
\affiliation
  {KEK Theory Center, Institute of Particle and Nuclear Studies, KEK,\\
   Oho 1-1, Tsukuba, Ibaraki, 305-0801, Japan}
\affiliation
{Institute of Modern Physics, Chinese Academy of Sciences, Lanzhou 730000, China}
\author{R. Kunitomo}
\affiliation
  {Department of Mathematics, Physics, and Computer Science,
   Faculty of Science, Japan Women's University,\\           
   Mejirodai 2-8-1, Bunkyo-ku, Tokyo 112-8681, Japan}
\author{Siyu Wu}
\affiliation
{Institute of Modern Physics, Chinese Academy of Sciences, Lanzhou 730000, China}
\affiliation
{South China Normal University, Guangzhou 510006, China}
\author{Ya-Ping Xie}
\affiliation
{Institute of Modern Physics, Chinese Academy of Sciences, Lanzhou 730000, China}
\affiliation
{University of Chinese Academy of Sciences, Beijing 100049, China}

\begin{abstract}

Spacelike and timelike generalized parton distributions (GPDs)
have been investigated at charged-lepton accelerator facilities, 
for example, by the virtual Compton scattering
and the two-photon process, respectively.
In this work, we show the $\pi^\pm$ and $\pi^0$ production cross sections
in neutrino (antineutrino) reactions $\nu \,(\bar\nu) + N \to \ell + \pi + N'$
for studying the GPDs of the nucleon by using
the theoretical formalism of Pire, Szymanowski, and Wagner.
The $\pi^\pm$-production cross sections are useful for finding mainly
gluon GPDs, whereas the $\pi^0$ production probes quark GPDs,
although there are sizable quark-gluon interference terms
in the $\pi^\pm$ production.
In particular, we show the roles of the pion- and rho-pole terms,
the contribution from each GPD term ($H$, $E$, $\tilde H$, $\tilde E$),
the effects of the gluon GPDs $H_g$ and $E_g$,
and the dependence on the energy of the process
in the neutrino cross sections.
These cross sections could be measured at Fermilab in future.
The neutrino GPD studies will play a complementary role
to the projects of charged-lepton and hadron reactions
for determining the accurate GPDs.

\end{abstract}
\maketitle

\section{Introduction}
\label{intro}

Understanding the origin of masses in nature 
is one of important issues as fundamental physics. 
The origin of hadron masses is not easily understood
because the hadrons are many-body systems of quarks and gluons,
and the masses are created by complicated emergent mechanisms
of their elementary constituents. The mass distributions in hadrons
are investigated by gravitational form factors (GFFs),
which provide key information for understanding the origin of 
the hadron masses. Because the gravitational interactions 
are too weak to measure the GFFs, in comparison with
the strong, electromagnetic, and weak interactions,
the GFFs used to be considered purely academic quantities
without any experimental measurement until several years ago.

The situation has changed because of the development
of hadron-tomography field in terms of 
generalized parton distributions (GPDs) 
\cite{gpds,partons-2018,mass-decomposition}.
The second moments of the GPDs are related
to the form factors of the energy-momentum tensor 
of quarks and gluon. These form factors are conventionally called 
the GFFs because a graviton couples to a quark or a gluon in the form 
of the energy-momentum tensor.
Alternatively, they are simply called the form factors
of the energy-momentum tensor.
Therefore, without relying on the direct gravitational
interactions, one can measure the GFFs through 
the electromagnetic interactions, or weak interactions
as shown in this paper.
In fact, gravitational radii were determined for the pion
by using the two-photon experimental data of the KEK-B factory
\cite{KEKB-2016,Kumano:2017lhr}. 
The two-photon process or the $s$-channel Compton scattering
is also possible by using the ultraperipheral hadron collisions 
\cite{Xie:2022vvl}.
Furthermore, the gravitational 
form factor and mass radius were also determined 
for the proton by using the vector-meson 
photo-production data \cite{vector-meson-proton-form}.
On the other hand, the renormalization of the quark and 
gluon energy-momentum tensors was investigated 
in perturbative quantum chromodynamics (QCD) \cite{pQCD-EMT}.
The GPDs also contain pressure and shear force information
in the hadrons \cite{nucleon-pressure,Kumano:2017lhr}, 
so their kinematical stability should be clarified.

The GPDs will provide not only the gravitational
information, such as the mass and pressure distributions, 
but also information on solving 
the origin of hadron spins by clarifying orbital angular 
momentum contributions \cite{gpds,mass-decomposition}.
The GPDs contain three dimensional quark and gluon distributions,
namely longitudinal parton distribution functions (PDFs)
and transverse form factors in the hadrons.
Because of this nature, the GPDs could be used also for clarifying
structure of exotic hadron candidates in future \cite{gpd-exotic}.

The spacelike GPDs have been investigated by
the deeply virtual Compton scattering (DVCS)
($\gamma^* +h \to \gamma +h$) where 
$\gamma$ is a photon and $h$ is a hadron,
meson ($M$) productions
($\gamma^* +h \to M +h$), and
J/$\psi$ production
($\gamma^* +h \to J/\psi +h$) \cite{vector-meson-proton-form}.
The generalized distribution amplitudes (GDAs) were studied
by the two-photon processes or $s$-channel Compton scattering
($\gamma^* + \gamma \to h + \bar h$)
\cite{KEKB-2016,Kumano:2017lhr}.
The GDAs could be called the $s$-channel GPDs because they are investigated
by the $s$-$t$ crossing process of the virtual Compton scattering,
where $s$ and $t$ are Mandelstam variables, and because the GDAs contain timelike form factors.
In addition, the GPDs could be investigated
in hadron reactions such as 
the exclusive Drell-Yan ($\pi + N \to \mu^+\mu^- +B$) \cite{J-PARC-gpds-exdy-2016}
and the hadronic $2 \to 3$ processes ($N + N \to N+ \pi +B$) 
\cite{kss-2009,J-PARC-GPDs,transition-GPDs}
where $B$ is a baryon.
There is also a recent theoretical proposal to investigate the GPDs
by the $2 \to 3$ reactions $\pi N \to \gamma\gamma N$,
$h M_A \to h \gamma M_B$, and $h M_A \to h M_B M_C$ 
where $M_{A,B,C}$ are mesons \cite{qiu-yu-2022}.
Furthermore, pseudoscalar-meson production processes 
could be studied for determining the GPDs 
at future electron-ion colliders (EICs) \cite{Goloskokov:2022mdn}.
The GPDs could be also investigated for hadrons with higher spins,
such as spin 1 like the deuteron \cite{spin-1}
and spin 3/2 like $\Delta$ and $\Omega$ \cite{spin-3/2}.

In this work, we show that the GPDs can be studied
in weak-interaction processes by showing neutrino cross
sections and their relations to the GPDs.
Neutrino reactions are complementary to charged-lepton
deep inelastic scattering experiments in determining
structure functions and parton distribution functions (PDFs)
of the nucleon. The neutrino reactions are sensitive
to the quark flavor ($e.g.$ $d+W^+ \to u$, $s+W^+ \to c$),
whereas the charged-lepton cross sections are 
proportional to quark-charge squared $e_q^2$.
In addition, the valence-quark distributions are determined
by the structure function $F_3$, and the strange-quark
distribution is found by the neutrino-induced 
opposite-sign dimuon production.

So far, there is no experimental project on
the GPDs in neutrino reactions.
However, as obvious in the unpolarized PDF determination,
the neutrino reactions should be valuable also
in the GPD studies. Fortunately, the high-energy neutrino
and antineutrino beams will become available at
the Long-Baseline Neutrino Facility (LBNF) of Fermilab
in the energy region of 2-15 GeV
\cite{lbnf-beam,kp-2021}, and it can be used
for the GPD and GFF studies.
In addition, there is a possibility that 
the neutrinos from stored muons (nuSTORM) projects \cite{nustorm,lu-2024} 
could have high-energy neutrino beam option for the GPD studies.

Theoretically, there were following cross section estimates
for studying the GPDs in the neutrino reactions.
In Ref.\,\cite{LS-2001}, $D_s$ production in neutrino scattering was investigated.
Deeply virtual neutrino scattering (DVNS) 
was formulated for the neutral current \cite{ACG-2005} and the charged current 
\cite{CG-2005}.
A detailed comprehensive analysis was given for the DVNS process for
the charge and neutral current amplitudes and cross sections
at the leading twist \cite{PMR-2007}.
Neutrino-induced pion production process was investigated in Ref.\,\cite{GHLM-2010}
as an extension of the pion electroproduction.
The $\pi$, $K$, and $\eta$ productions in neutrino scattering were studied
at the leading twist and the leading order \cite{KSS-2012}.
Twist-3 contributions to the pion production were investigated
with the chiral-odd transversity GPDs \cite{KSS-2014}.
By the $D$-meson production, the transversity chiral-odd GPDs
could be found by the cross sections up to $m_c/Q$ \cite{PS-2015}.
NLO corrections to the pion and kaon productions were investigated
in Ref.\,\cite{SS-2017}. 
The pseudoscalar charmed-meson production was studied in the leading order 
by including the gluon GPDs \cite{psw-2017-charm}, whereas previous studies were only 
on the quark GPDs. The most updated information was given in
the formalism of Pire, Szymanowski, and Wagner 
in 2017 (PSW-2017) \cite{psw-2017} for productions of light mesons
by including gluon GPD contributions.

We calculate the neutrino cross sections 
by using the formalism of the PSW-2017 
and try to show
\vspace{-0.10cm}
\begin{enumerate}
\setlength{\leftskip}{+0.40cm}
\setlength{\itemsep}{-0.05cm} 
\item 
contribution from each GPD ($H$, $E$, $\tilde H$, $\tilde E$),
\item 
effects of pion- and rho-meson pole GPDs,
\item 
effects of the gluon GPDs $H_g$ and $E_g$,
\item 
dependence on the energy of the process,
\vspace{-0.10cm}
\end{enumerate}
in the cross sections.
The meson-pole terms exist 
in the Efremov-Radyushkin-Brodsky-Lepage (ERBL) region, 
which is not experimentally investigated so far
\cite{kss-2009,J-PARC-GPDs,transition-GPDs}. 
It is valuable to show the effects of these undetermined GPDs
in the neutrino and antineutrino cross sections.
Because the unpolarized gluon distribution is determined
relatively well from various experiments, there are some
constraints on the gluon GPD $H_g$. However, the gluon GPD $E_g$
is not determined at this stage.
Since the LBNF experiment is under preparation 
for the GPD experiment, it is valuable to show
the reliable cross sections numerically and 
to clarify what kind of the GPDs can be
studied in the neutrino reactions in comparison
with the charged-lepton GPD studies at 
the Thomas Jefferson National Accelerator Facility (JLab),
CERN-AMBER, KEK-B and future EICs \cite{kunitomo-2023}.

This paper consists of the following.
In Sec.\,\ref{cross}, the cross section formalism is shown 
with a brief introduction to the GPDs
and pion- and rho-distribution amplitudes.
Cross section results are shown in Sec.\,\ref{results}
for the $\pi^+$ and $\pi^0$ productions in neutrino reactions.

\section{Neutrino-nucleon cross sections and GPDs}
\label{cross}

\subsection{Generalized parton distributions}
\label{gpds}

The GPDs are investigated in the neutrino reaction
$W^* (q) + N (p) \to \pi (q') + N (p')$, 
and they are expressed by the three variables, 
the Bjorken variable $x$, the skewness parameter $\xi$,
and the momentum-transfer squared $t$. 
These variables are expressed by 
the initial (final) nucleon, $W$, and pion momenta 
$p$ ($p'$), $q$, and $q'$, respectively.
The average momenta $\bar P$ and $\bar q$
and the momentum transfer $\Delta$ are defined by
\begin{align}
\! \! \! 
\bar P = (p+p')/2,  \ 
\bar q = (q+q')/2,  \  
\Delta = p'-p = q-q'.
\label{eqn:pqd}
\end{align}
Then, $Q^2$, $\bar Q^2$ and the GPD variables $x$, $\xi$, and $t$ 
are expressed as
\begin{alignat}{2}
Q^2 & = -q^2 , \ \ &
\bar Q^2 & = - \bar q^2 , 
\nonumber \\
x & = \frac{Q^2}{2p \cdot q} , \ \ &
\xi & = \frac{\bar Q^2}{2 \bar P \cdot \bar q}, \ \ 
t = \Delta^2 .
\label{eqn:gpd-variables}
\end{alignat}
For the reaction with $Q^2 \gg |t|$, 
the skewness parameter 
is directly related to the variable $x$ as 
\cite{gpd-exotic}
\begin{align}
\xi = \frac{x \, [ 1 + t / (2 Q^2) ]}{2 - x \, ( 1 - t / Q^2 ) } 
    \simeq \frac{x}{2 - x} \ \  \text{for} \ Q^2 \gg |t| .
\label{eqn:xi-x}
\end{align}
Because this relation is generally satisfied 
in considered neutrino reactions of this work,
it is used in our numerical analysis.

\begin{figure}[b!]
 \vspace{-0.30cm}
\begin{center}
   \includegraphics[width=6.5cm]{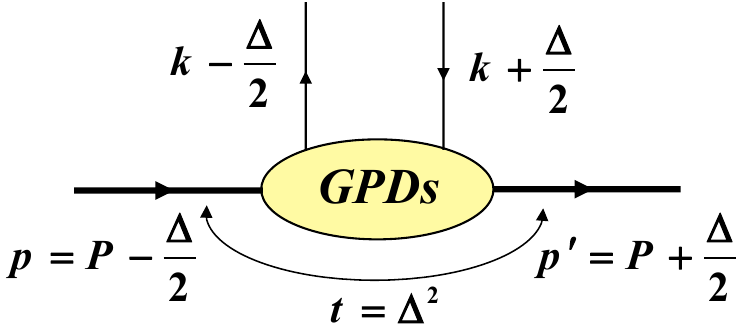}
\end{center}
\vspace{-0.5cm}
\caption{GPD kinematics.}
\label{fig:gpd-kinematics}
\vspace{-0.00cm}
\end{figure}

These quantities are expressed by the lightcone variables
$a=(a^+, \, a^-, \, \vec a_\perp)$.
The variables $a^\pm$ are defined by 
$a^\pm = (a^0 \pm a^3)/\sqrt{2}$, and $\vec a_\perp$ is
the transverse vector. 
By neglecting the nucleon mass, 
the nucleon and photon momenta are expressed as
\begin{alignat}{2}
& \! \! 
p \simeq \left ( p^+, \, 0, \, \vec 0_\perp \right ) , \ \ \, &
& p' \simeq \left ( {p'}^+, \, 0, \, \vec 0_\perp \right ) ,
\nonumber \\
& \! \! 
q \simeq \left ( -x p^+, \, \frac{Q^2}{2xp^+}, \, \vec 0_\perp \right ) , \ \ \, &
& q' \simeq \left ( 0,     \,  \frac{Q^2}{2xp^+}, \, \vec 0_\perp \right ) ,
\label{pandq}
\end{alignat}
where the relation $(p^+)^2 , \, Q^2 \gg m_N^2, \, |t|$ is assumed,
and the momentum conservation indicates ${p'}^+ \simeq (1-x) p^+$.
Here, $m_N$ is the nucleon mass.
Then, the variable $x$ and $\xi$ are written
by the lightcone variables as
\begin{align}
x \simeq \frac{k^+}{P^+}, \ \ \ 
\xi \simeq -\frac{\Delta^+}{P^+} ,
\label{eqn:xi-x-lightcone}
\end{align}
with the total momentum $P=p+p'$ and the average parton momentum $k$
between the initial and final partons,
as shown in Fig.\,\ref{fig:gpd-kinematics}, in the nucleon.
The scaling variable $x$ is the momentum fraction carried by a parton,
and the skewness parameter $\xi$ is the momentum-transfer 
fraction from the initial nucleon to the final nucleon or 
from the initial quark to the final quark.

There are three kinematical regions in the GPDs
depending on the scaling variable $x$ and a skewness parameter $\xi$:
\begin{itemize}
\vspace{-0.10cm}
\setlength{\itemsep}{-0.05cm} 
\setlength{\leftskip}{0.75cm}
\item[$(a)$] $-1 < x < -\xi$:\,antiquark distribution,
\item[$(b)$] $-\xi < x <\xi$: \hspace{0.05cm} meson\,(quark-antiquark)\,distribution,
\item[$(c)$] $\xi < x <1$:    \hspace{0.30cm} quark distribution.
\vspace{-0.10cm}
\end{itemize}
\noindent
The regions $(a)$ and $(c)$ are called the
Dokshitzer-Gribov-Lipatov-Altarelli-Parisi (DGLAP) regions, 
and the region $(b)$ is called the ERBL region.
All of these regions need to be understood for clarifying
the GPD functions.

The GPDs are defined by the off-forward matrix element of
bilocal quark (and gluon) operators 
with a light-like separation as
\begin{align}
&  \int \frac{d y^-}{4\pi}e^{i x P^+ y^-}
 \langle p' \left| 
 \bar{q}(-y/2) \gamma^+ q(y/2) 
 \right| p \rangle
  _{y^+ = \vec y_\perp =0}
\nonumber \\
& = 
 \frac{1}{2  P^+} \bar{u} (p') 
  \left [ H^q (x,\xi,t) \gamma^+
  + E^q (x,\xi,t)  \frac{i \sigma^{+ \alpha} \Delta_\alpha}{2 m_N}
 \right ] u (p) ,
\label{eqn:gpd-vector}
\nonumber\\[-0.25cm]
\end{align}
\ \vspace{-1.00cm}
\begin{align}
& \int \frac{d y^-}{4\pi}e^{i x P^+ y^-}
 \langle p' \left|  \bar{q}(-y/2) \gamma^+ \gamma_5 q(y/2) 
 \right| p \rangle
 _{y^+ = \vec y_\perp =0}
\nonumber \\
& = \frac{1}{2  P^+} \bar{u} (p') 
\left [ \tilde{H}^q (x,\xi,t) \gamma^+ \gamma_5
    +  \tilde{E}^q (x,\xi,t)  \frac{\gamma_5 \Delta^+}{2 m_N}
 \right ] u (p) ,
\label{eqn:gpd-axial-v}
\nonumber\\[-0.20cm]
\end{align}
where $u(p)$ denote the Dirac spinor for the nucleon,
and $\sigma^{\alpha\beta}$ is defined by
$\sigma^{\alpha\beta}=(i/2)[\gamma^\alpha, \gamma^\beta]$.  
In these equations, we abbreviated the gauge-link operator between
the two quark fields for the color gauge invariance. 
The functions $H^q (x,\xi,t)$ and $E^q (x,\xi,t)$ are the unpolarized quark GPDs, 
and $\tilde{H}^q (x,\xi,t)$ and $\tilde{E}^q (x,\xi,t)$ are the polarized ones. 
We also do not write the scale dependence of the GPDs 
from that of the bilocal operators explicitly.

Three important features of the GPDs are the following.
First, the functions $H^q (x,\xi,t)$ and $\tilde{H}^q (x,\xi,t)$ 
become the unpolarized and longitudinally-polarized PDFs
in the forward limit
\begin{align}
H^q (x, 0, 0) = q(x), \ \ 
\tilde{H}^q (x, 0, 0) = \Delta q(x) .
\end{align}
Second, the first moments of $H^q (x,\xi,t)$, $E^q (x,\xi,t)$,
$\tilde{H}^q (x,\xi,t)$, and $\tilde{E}^q (x,\xi,t)$ 
are the Dirac, Pauli, axial and pseudoscalar form factors, respectively
\begin{align}
\ \hspace{-0.30cm}
\int_{-1}^{1} dx H^q(x,\xi,t)  & = F_1^q (t), \ 
\int_{-1}^{1} dx E^q(x,\xi,t)    = F_2^q (t),
\nonumber\\
\ \hspace{-0.30cm}
\int_{-1}^{1} dx \tilde{H}^q(x,\xi,t)  & = g_A^q (t),  \ 
\int_{-1}^{1} dx \tilde{E}^q(x,\xi,t)    = g_P^q (t) .
\end{align}
Third, the second moments of 
$H^q (x,\xi,t=0)+E^q (x,\xi,t=0)$
indicate the contributions from the quark angular momenta
\begin{align}
J^q & = \frac{1}{2} \int_{-1}^{1} dx \, x \, [
       H^q (x,\xi,t=0) +E^q (x,\xi,t=0) ] 
\nonumber\\
    & = \frac{1}{2} \Delta q + L^q, 
\end{align}  
so that the quark orbital-angular-momentum contribution $\sum_q L^q$
should become clear to the nucleon spin.

The nucleon mass is defined by the matrix element 
\cite{mass-decomposition}
\begin{align}
m_N = \left \langle \, N(p) \, \left | \, \int d^3r \, T^{00} (\vec r \,) \, 
         \right | \, N(p) \, \right \rangle ,
\label{eqn:n-mass}
\end{align}
where the static energy-momentum tensor $T^{\mu\nu} (\vec r \,)$
is defined in the Breit frame as
\begin{align}
\! \! \!
T^{\mu\nu} (\vec r \,) = \int \frac{d^3 q}{(2\pi)^3 2E} e^{i \vec q \cdot \vec r}
     \left \langle \, N(p') \, \left | \,T^{\mu\nu} (0) \, 
         \right | \, N(p) \, \right \rangle ,
\label{eqn:Tmunu}
\end{align}
with $E = \sqrt{m_N^2 + \vec q^{\,2}/4}$, and
the energy-momentum tensor $T^{\mu\nu}$ on the right-hand side 
is given by the quark and gluon fields as
\begin{align}
    T^{\mu\nu}(x) & =\,\bar q (x)\,\gamma^{(\mu} i \overleftrightarrow{D}^{\nu )} \,q(x)
\nonumber \\
   & \ \ \ 
    +\left [ \frac{1}{4}\,g^{\mu\nu}\,F^2(x)\,
    -\,F^{\mu\alpha}(x)\,F_\alpha^\nu(x)
  \right ]
\nonumber\\
&  \equiv \,T_q^{\mu\nu}(x)\,+\,T_g^{\mu\nu}(x) .
\label{eqn:tmunu}
\end{align}
Here, $D$ is the covariant derivative defined by
$D_\mu  = \partial_\mu  - i g T^a A_\mu^{a} (x)$
with the QCD coupling constant $g$
and the gluon field 
$A_\mu = \sum_{a=1}^{N^2-1} T^a A_\mu^{a} \ (\equiv  T^a A_\mu^{a})$ 
where $T^a$ is expressed by the Gell-Mann matrix $\lambda^a$
as $T^a=\lambda^a /2$.
The derivative $\overleftrightarrow{D}$ is 
$f_1 \overleftrightarrow{D} f_2 = f_1 (D f_2) - (D f_1) f_2$.
The first term of the right-hand side in Eq.\,(\ref{eqn:tmunu}) 
is symmetrized under the indices as
$A^{(\mu} B^{\nu )} = (A^{\mu} B^{\nu}+ A^{\nu} B^{\mu})/2$.
The gluon field strength $F_{\mu\nu}^{\,a}(x)$ is
$ F_{\mu\nu}^{\,a} (x)
= \partial_\mu \, A_\nu^a (x) - \partial_\nu \, A_\mu^a (x) 
 + g \, f^{\, abc} \, A_\mu^b (x) \, A_\nu^c (x) $, where
$f^{\, abc}$ the structure constant of the $SU(3)$ group.
In Eq.\,(\ref{eqn:tmunu}), the gluon field strength $F$ is
$F^{\mu\nu}=F_a^{\mu\nu} T_a$, and 
$F^2$ is $F^2 = F^{\mu\nu} F_{\mu\nu}$.

We write the matrix element of the energy-momentum tensor
in terms of the gravitational form factors 
$A$, $B$, $\bar{C}$, and $D$ of the nucleon as \cite{nucleon-pressure}
\begin{align}
    \langle \,N(p^\prime)\,| & \,T^{\mu\nu}_{q,g} (0)\,|\,N(p)\,\rangle
     = \bar u(p^\prime)\, \biggl[
      A_{q,g} \gamma^{(\mu} \bar P^{\nu)}
\nonumber\\
   & 
    + B_{q,g} \frac{\bar P^{(\mu} i \sigma^{\nu)\alpha} \Delta_\alpha}{2m_N}
    + D_{q,g} \frac{\Delta^\mu\Delta^\nu-g^{\mu\nu}\Delta^2}{m_N}
\nonumber\\
   & 
    +\bar{C}_{q,g} m_N g^{\mu\nu}
     \biggr]\,u(p) .
    \label{Eq:EMT-FFs-spin-12} 
\end{align}
The form factor $A$ indicates the mass (energy) distribution,
$B$ does the angular momentum distribution,
$D$ does the pressure distribution,
and $\bar C$ does the mass and pressure distribution.
Its $\mu\nu=00$ component corresponds to the mass distribution in the nucleon:
\begin{align}
& \! \! 
\langle \,N(p^\prime)\,| \,T_{q,g}^{00}(0)\,|\,N(p)\,\rangle
    = 2 m_N E\,\biggl[ A_{q,g}(t) 
\nonumber \\
      &
        - \frac{t}{4 m_N^2} \bigl\{ A_{q,g}(t)-2B_{q,g}(t)+D_{q,g}(t)\bigr\}
        + \bar C_{q,g} (t)	\biggr]  .
	\label{Eq:EMT-T00}
\end{align}
These gravitational form factors are expressed by the GPDs as
\begin{align}
	\int_{-1}^1{\rm d}x\;x\, H^{q,g} (x,\xi,t) 
	    & = A_{q,g}(t) + 4 \xi^2 D_{q,g}(t) , 
\nonumber\\
        \int_{-1}^1{\rm d}x\;x\, E^{q,g} (x,\xi,t) 
        & = B_{q,g}(t) - 4 \xi^2 D_{q,g}(t) .
\label{Eq:GPD-2}
\end{align}
If these GPDs of the nucleon are determined, it leads to the understanding
to the origin of the nucleon mass in terms of quarks and gluons.

\subsection{Pion and rho distribution amplitudes \\ and their-pole GPDs}
\label{pion-rho-da}

For describing the pion-production cross section, 
the pion distribution amplitude is necessary.
In addition, we investigate effects of pion- and rho-pole terms
in the nucleon GPDs on the cross sections. In these pole terms,
the pion- and rho-distribution amplitudes exist. Therefore,
their distribution amplitudes and pole GPDs are explained
in this subsection.

The twist-2 pion distribution amplitude $\Phi_\pi (x,\mu)$
is defined by the matrix element between the vacuum and pion states as
\cite{mueller-1989,cz-1984,J-PARC-gpds-exdy-2016}
\begin{align}
& \langle \, 0 \, |   \, \bar d (0)_\alpha \, 
                  \, u (y)_\beta \, 
                | \, \pi^+ (p_\pi) \, \rangle\Big |_{y^- = \vec y_\perp =0}
\nonumber \\
& \ \ 
= \frac{i f_\pi}{4} \int_0^1 dx \, e^{-i x p_\pi^- y^+}
  \big ( \gamma_5 \, \pslash_\pi \big )_{\beta\alpha} \, \Phi_\pi (x,\mu) ,
\label{eqn:pion-da-x}
\end{align}
where higher-twist terms are abbreviated on the right-hand side.
The variable $x$ indicates the momentum fraction of a valence quark
in the pion, and $\mu$ is the renormalization scale of the bilocal
operator on the left-hand side of Eq.\,(\ref{eqn:pion-da-x}).
The pion decay constant $f_\pi$ is define by
$\langle \, 0 \, | \, \bar d (0) \gamma^\mu
\gamma_5 \, u(0) \, | \, \pi^+ (p_\pi) \, \rangle = i f_\pi
p_\pi^\mu$, and the pion distribution amplitude $\Phi_\pi (x,\mu)$
satisfies the normalization
$\int_0^1 dx \, \Phi_\pi (x, \mu)=1$.

The pion distribution amplitude is often used
in the kinematical region $0<x<1$. 
However, the range of the variable $x$ is between $-1$ and $1$
in the GPDs as explained in the previous subsection.
The pion distribution amplitude needs to be converted 
by using this variable. In order to avoid confusion, 
we express this variable between $-1$ and $1$ as $z$
only in this subsection.
Using the relation $z=2x-1$, we rewrite Eq.\.(\ref{eqn:pion-da-x}) as
\begin{align}
& \langle \, 0 \, | \, \bar d (-y)_\alpha \, 
                  \, u (y)_\beta \,
                | \, \pi^+ (p_\pi) \, \rangle\Big |_{y^- = \vec y_\perp =0}
\nonumber \\
& \ \ 
= \frac{i f_\pi}{4} \int_{-1}^1 dz \, e^{-i z p_\pi^- y^+}
  \big ( \gamma_5 \, \pslash_\pi \big )_{\beta\alpha} \, \phi_\pi (z,\mu) ,
\label{eqn:pion-da-z}
\end{align}
where the pion distribution amplitude $\phi_\pi (z,\mu)$ is 
expressed by $\Phi_\pi (x,\mu)$ as
\begin{align}
& \phi_\pi (z,\mu)=\frac{1}{2}\Phi_\pi\left(x= \frac{z+1}{2},\mu\right), 
\nonumber \\
& \ \ \ \ 
\phi_\pi (z,\mu): \ \ \ -1 < z < 1,
\nonumber \\
& \ \ \ \ 
\Phi_\pi (x,\mu):  \ \ \ 0 < x < 1,
\label{eqn:pion-da-relation}
\end{align}
with the normalization $\int_{-1}^1 dz \, \phi_\pi (z, \mu)=1$.
This relation (\ref{eqn:pion-da-relation}) should be noted
in discussing the pion- and rho-pole terms in the GPDs.

The pion state is written by the Fock states 
in the lightcone quantization as 
\begin{equation}
|\pi(p_\pi)\rangle =\int \frac{dx \, d^2 \vec k_T}{16\pi^3\sqrt{x\bar{x}}}
\Psi_{q\bar{q}/\pi}(x, \vec k_T)
|q(k_q)\bar{q}(k_{\bar{q}})\rangle+\cdots ,
\label{eqn:pion-light-cone}
\end{equation} 
where $\Psi_{q\bar{q}/\pi}(x, \vec k_T)$ is 
the Bethe-Salpeter (BS) wave function, 
$|q(k_q)\bar{q}(k_{\bar{q}})\rangle$ is the $q\bar{q}$ Fock state,
and the ellipsis ($\cdots$) indicates higher Fock states.
The variables $x$ and $\bar x (=1-x)$ are momentum fractions
of the quark and antiquark, respectively, and $\vec k_T$
is the quark transverse momentum
($\vec k_T \equiv \vec{k}_{q T}=-\vec{k}_{\bar{q} T}$).
The pion distribution amplitude is given by this BS wave function as
\cite{mueller-1989}
\begin{equation}
\Phi_\pi(x,\mu) = \frac{2\sqrt{2 N_c}}{if_\pi}
\int_{|\vec k_T|<\mu} \frac{d^2 \vec k_T}{16\pi^3}
\Psi_{u\bar{d}/\pi}(x,\vec k_T) ,
\label{eqn:pion-DA-BS}   
\end{equation}
with the number of colors $N_c$.

In general, the distribution amplitude is expressed by 
the Gegenbauer polynomials $C_n^{3/2}$ at $\mu$ as
\begin{equation}
\Phi_\pi (x,\mu) = 6 \, x \, (1-x) 
\sum_{n=0,2,4,\cdots}^\infty a_n (\mu) \, C_n^{3/2} (2x-1).
\label{eqn:gegenbauer}
\end{equation}
In the asymptotic limit $\mu \to \infty$ (as), it becomes
\cite{erbl-1979} 
\begin{equation}
\Phi_\pi^{\text{as}} (x) = 6 \, x \, (1-x) ,
\label{eqn:asym-pion}
\end{equation}
whereas the following form with $a_2\,(\mu\simeq 0.5~{\rm GeV}) = 2/3$
was proposed by Chernyak and Zhitnitsky (CZ) \cite{cz-1984}
\begin{align}
\Phi_\pi^{\text{CZ}} (x, \mu\simeq 0.5~{\rm GeV}) 
& = 30 \, x \, (1-x) \, (2x-1)^2 .
\label{eqn:cz}
\end{align}
For calculating the neutrino cross section in Sec.\,\ref{results},
the pion distribution amplitude of Ref.\,\cite{Kroll-2013} is used
with $a_2=0.22$ and $a_{4,6,\cdots}=0$
in Eq.\,(\ref{eqn:gegenbauer}).
The rho distribution amplitude $\Phi_\rho (x,\mu)$ 
is also given in the same way \cite{rho-da},
and its asymptotic form is given as 
\begin{equation}
\Phi_\rho^{\text{as}} (x) = 6 \, x \, (1-x) .
\label{eqn:asym-rho}
\end{equation}

Using these distribution amplitudes, we write the pion- and rho-pole
contributions to the GPDs.
The pion- and rho-pole GPDs are given in Ref.\,\cite{kss-2009}, and 
different normalizations are used for their distribution amplitudes
from the ones explained in this section.
The pion decay constant, pion distribution amplitude, and 
rho distribution amplitude of Ref.\,\cite{kss-2009} are
denoted $f'_\pi$, $\Phi'_\pi$, and $\Phi'_\rho$ to avoid confusion.
Then, they are related to $f_\pi$, $\Phi_\pi$, and $\Phi_\rho$
in this section as
\begin{align}
& f'_\pi = \frac{1}{\sqrt{2}} f_\pi, 
\nonumber \\
& \Phi'_\pi (x) = \frac{f_\pi}{2\sqrt{6}} \Phi_\pi (x), \ \ 
  \Phi'_\rho (x) = \frac{f_\rho}{\sqrt{6}} \Phi_\rho (x),
\end{align}
with the rho decay constant $f_\rho$, so that
the Goldberger-Treiman relation is given as
$f'_\pi g_{\pi NN} = m_N g_A$ with the pion-nucleon coupling 
constant $g_{\pi NN}$ and 
the axial charge of the nucleon $g_A$.

The pion is a pseudo-scalar hadron and it contributes to 
the axial GPD $\tilde E (z,\xi,t)$, 
where $z$ is used for indicating the range $-1 <z<1$ explicitly
in this subsection, as \cite{gpv01,ppg00,kss-2009}
\begin{align}
& \! \! \!
\tilde E_\pi (z,\xi,t)
  = \frac{4 \, g_A \, m_N^2}{1-t/m_\pi^2} 
  \frac{\sqrt{3}}{m_\pi^2 \, f'_\pi \, \xi} 
    \, \Phi'_\pi \left (x= \frac{z+\xi}{2 \xi} \right)
     \theta(\xi -|z|) 
\nonumber \\
& \! \! \! \!
  = -m_N f_\pi \frac{2 \sqrt{2} g_{\pi NN}}{t-m_\pi^2} \frac{1}{2 \xi}
    \, \Phi_\pi \left (x= \frac{z+\xi}{2 \xi} \right)
    \theta(\xi -|z|) .
\label{eqn:pion-pole}
\end{align}
The function $\tilde E_\pi$ indicates the isovector combination
in terms of up and down quark GPDs as
$(\tilde E^u-\tilde E^d)_\pi$.
Therefore, each quark GPD is expressed as
\begin{align}
\tilde E^u (z,\xi,t) = - \tilde E^d (z,\xi,t)
= \frac{1}{2} \tilde E_\pi (z,\xi,t) .
\label{eqn:pion-pole-eqch-quark}
\end{align}
There is no contribution to the GPD $\tilde H$ from the pion.

The rho is a vector hadron, and it contributes
to the vector part.
The rho-pole GPDs are given by \cite{feynman,kss-2009}
\begin{align}
H_\rho (z,\xi,t)
& = {1\over 1-t/m_{\rho}^2} 
    \, \frac{C_1}{g_\rho}
     \frac{1}{2 \xi} 
    \, \Phi_\rho \left (x= \frac{z+\xi}{2 \xi} \right)
     \theta(\xi -|z|) ,
\nonumber \\
E_\rho (z,\xi,t)
& = {1\over 1-t/m_{\rho}^2} 
    \, \frac{C_2}{g_\rho}
      \frac{1}{2 \xi} 
     \, \Phi_\rho \left (x= \frac{z+\xi}{2 \xi} \right)
      \theta(\xi -|z|) ,
\label{eqn:rho-pole}
\end{align}
where $g_\rho$ is the vector coupling constant of the rho,
and $m_\rho$ is the rho mass.
The $C_1$ and $C_2$ are isovector rho coupling constants
with the nucleon, and they are given by
\begin{align}
\frac{C_1}{g_\rho}  =\frac{e_p-e_n}{2} ,
\ \ 
\frac{C_2}{g_\rho}  =
\frac{\kappa_p-\kappa_n}{2}
\end{align}
with the charge of the proton (neutron) $e_p$ ($e_n$),
and $\kappa_p$ ($\kappa_n$) is the anomalous magnetic
moment of the proton (neutron).
Then, the quark GPDs are given as
\begin{align}
H^u (z,\xi,t) & = - H^d (z,\xi,t) = \frac{1}{2} H_\rho (z,\xi,t),
\nonumber \\
E^u (z,\xi,t) & = -E^d (z,\xi,t) = \frac{1}{2} E_\rho (z,\xi,t) .
\label{eqn:rho-pole-eqch-quark}
\end{align}
These meson contributions to the GPDs exist 
in the ERBL region and they should be included
within the nucleonic GPDs.

\subsection{Cross section for the pion production \\ in neutrino reactions}
\label{pion-cross-section}

As the most recent theoretical formalism for describing
the neutrino-induced pion production 
($\nu N \to \ell N'\pi$) cross section
by using the GPDs, we use the PSW-2017 \cite{psw-2017}.
It includes not only quark contributions but also 
gluon ones for the $\pi^\pm$ and $\pi^0$ productions.
Since it is a weak-interaction process, it is appropriate
to rely on larger cross sections with the pion production experimentally,
instead of the photon production in the DVCS.

\begin{figure}[t!]
 \vspace{-0.00cm}
\begin{center}
   \includegraphics[width=4.5cm]{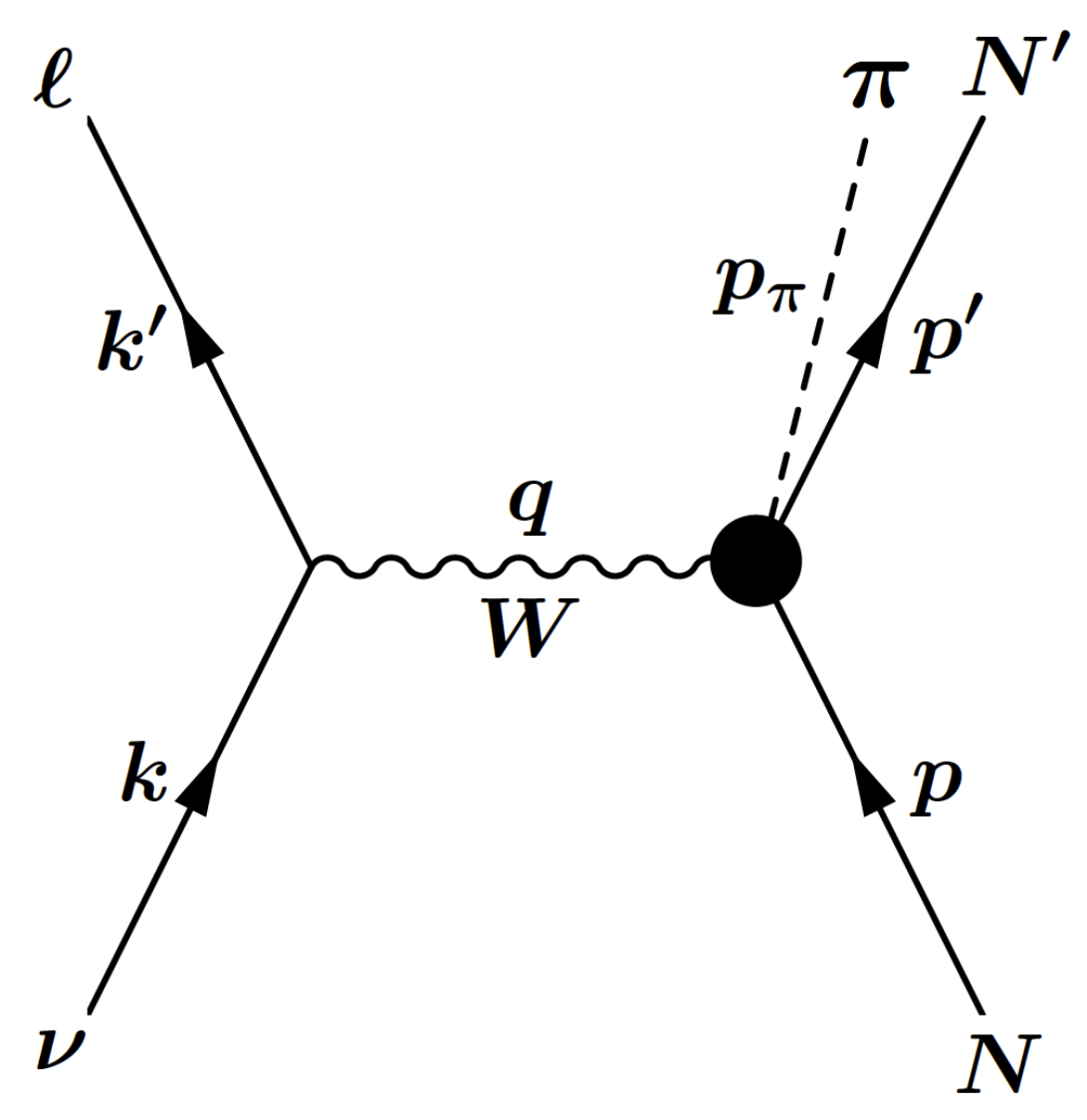}
\end{center}
\vspace{-0.5cm}
\caption{
Pion-production process $\nu N \to \ell N'\pi$
in a charged-current neutrino reaction.}
\label{fig:neutrino-pi}
\vspace{-0.30cm}
\end{figure}

\begin{figure}[b!]
 \vspace{-0.30cm}
 \hspace{0.05cm}
   \includegraphics[width=4.1cm]{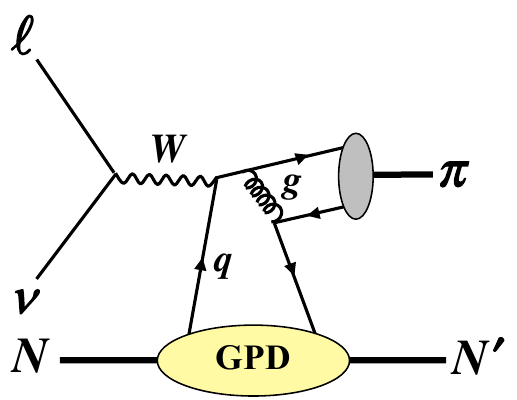}
 \hspace{0.05cm}
   \includegraphics[width=4.1cm]{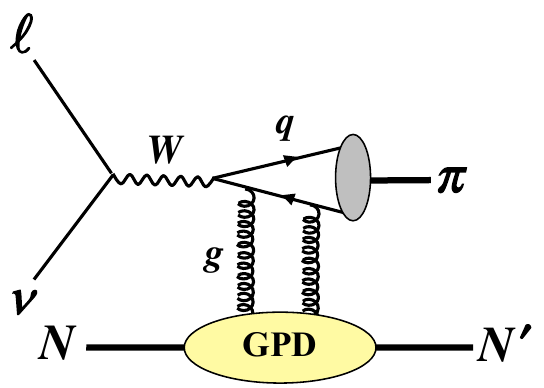}
\vspace{-0.2cm}
\caption{Typical pion-production 
processes with quark and gluon GPDs.}
\label{fig:neutrino-pi-gpd}
\vspace{-0.00cm}
\end{figure}

The pion-production process $\nu N \to \ell N'\pi$
is shown in Fig.\,\ref{fig:neutrino-pi}, where
$k$ ($k'$), $p$ ($p'$), $q$, and $p_\pi$ are the momenta 
for the initial neutrino (final lepton), 
the initial (final) nucleon, the W boson, and the pion, respectively.
This pion-production cross section is described 
in a factorized form with the GPDs as typically shown 
in Fig.\,\ref{fig:neutrino-pi-gpd}, where the quark and gluon GPDs
contribute to the pion production.
In describing its cross section, the kinematical variables $Q^2$, $x$, $t$,
$\Delta$ are defined in Eqs.\,(\ref{eqn:pqd}) and (\ref{eqn:gpd-variables}).
In addition, $y$ is given by $y= p \cdot q /p \cdot k$.
The momentum-transfer squared $Q^2$ is expressed by 
the center-of-mass-energy squared $s=(p+k)^2$ as $Q^2 = x y (s-m_N^2)$.

In general, both quark and gluon GPDs contribute to the pion-production
cross section, and their amplitudes $T^q$ and $T^g$ are given by
\cite{psw-2017}
\begin{align}
T^q  & = - i \, \frac{C_q}{2Q} \, \bar{u}(p') 
   \bigg(  {\mathcal{H}}^{q} \slashed{n} 
           - \tilde {\mathcal{H}} ^{q} \slashed{n} \gamma^5 
\nonumber \\
&  \hspace{2.3cm}
+ {\mathcal E}^{q} \frac{i\sigma^{n\Delta}}{2m_N} 
        - \tilde {\mathcal E}^{q} 
        \frac{\gamma ^5 \Delta \cdot n}{2m_N}\bigg) u(p),
\nonumber \\[-0.00cm]
T^g & =  - i \, \frac{C_g}{2Q}  \, \bar{u}(p') \left( {\cal H}^g \slashed{n}
      +{\cal E}^g\frac{i\sigma^{n\Delta}}{2m_N} \right) u(p) .
\label{eqn:quark-gluon-GPDs}
\end{align}
Here, the constants $C_q$ and $C_g$ are defined by
color and flavor factors as
$C_q = 2\pi C_F \alpha_s V_{du}/3$ with 
$C_F= (N_c^2 -1)/(2 N_c) = 4/3$ where the number of color is $N_c=3$, 
and $C_g = \pi T_f \alpha_s V_{du}/3$ with $T_f =1/2$.
The $\alpha_s$ is the running coupling constant of QCD,
and $V_{du}$ is the quark mixing angle, 
$Q$ is given by $Q= \sqrt{Q^2}$,
$\slashed{n}$ is given by $\slashed{n} \equiv n^\mu \gamma_\mu = \gamma^+$
with $n^\mu = (1,\,0,\,0,\,-1)/\sqrt{2}$,
and $\sigma^{n\Delta}$ is $\sigma^{n\Delta}=\sigma^{\mu\nu}n_\mu \Delta_\nu$.
The functions ${\cal F }^{q}$ and ${\cal F }^{g}$ 
(${\cal F}={\cal H}$, ${\cal \tilde H}$, ${\cal E}$, ${\cal \tilde E}$)
are defined by the integrals as
\vspace{-0.10cm}
\begin{align}
{\cal F }^{q} & = 2 f_{\pi}\int_0^1 dx' \frac{\Phi_\pi (x')}{1-x'}
    \int_{-1}^1 dx \frac{F^{q}(x,\xi,t)}{x-\xi +i\epsilon} , 
\nonumber \\
{\cal F }^{g} & = \frac{8 f_{\pi}}{\xi} 
    \int _0^1 dx' \frac{\Phi_\pi (x')}{x' (1- x')} 
\int _{-1}^{1}dx \frac{F^{g}(x,\xi,t)}{x-\xi +i\epsilon} ,
\label{eqn:quark-gluon-GPD-integral}
\end{align} 
where $\Phi_\pi (x')$ is the pion distribution amplitude 
defined in the range $0<x'<1$, 
and $F^{q}$ and $F^{g}$ are quark and gluon GPDs
($F=H$, $E$, $\tilde H$, $\tilde E$).

The quark GPD contribution is shown on the left-hand side
of Fig.\,\ref{fig:neutrino-pi-gpd}, where the quark $q$ 
could be up, down, strange, or charm quark.
In our numerical analyses, the strange- and charm-quark contributions 
are neglected, and the up and down quark
GPDs $F^u$ and $F^d$ contributions are estimated.
If the intermediate W-boson is $W^+$ ($\nu p \to \ell^- W^+ p$),
a $d$-quark is extracted from the proton to interacts with $W^+$
and it becomes a $u$-quark to form a $\pi^+$
together with a $\bar d$ quark. Then, a $d$-quark is absorbed
into the proton to form a proton in the final state.
This amplitude is described by the GPD $F^d$.
In the same way, the $W^+$ boson interacts with 
an antiquark $\bar u$ from the initial proton.
According to Table 1 in the paper of 
Kopeliovich-Schmidt-Siddikov of 2012 \cite{KSS-2012},
these two process amplitudes are
given by 
$ V_{ud} [H_d(x) c_- + H_u(x) c_+] \to 
  V_{ud} \int dx [H_d(x)/(x-\xi+i \epsilon) 
                +H_u(x)/(x+\xi-i \epsilon)]
= V_{ud} \int dx [ H_d(x) - H_u(-x) ]/(x-\xi+i \epsilon) $,
where $c_\pm = 1/(x\pm\xi\mp i\epsilon)$.
Therefore, the charged-current reactions are sensitive 
to the $u$- and $d$-quark GPDs by the following combination
\cite{psw-2017}
\begin{align}
& F^{q}(x,\xi,t) = F^d(x,\xi,t) - F^u(-x,\xi,t)
\nonumber \\
& \ \ \ \ 
\text{for the $\pi^+$ production $\nu  p \to \ell^- p \, \pi^+$}.
\label{eqn:pi+-production-quark-GPD}
\end{align}
On the other hand, all the $u$, $d$, $\bar u$, and $\bar d$ quarks
contribute to the $\pi^0$-production cross section 
($\bar\nu p \to \ell^+ n \, \pi^0$), 
and its amplitude is given by 
$ V_{ud} [H_u(x) - H_d(x)](c_+ - c_-)/\sqrt{2} $ 
\cite{KSS-2012}.
For the $\pi^0$ production with the initial neutron
($\nu n \to \ell^- p \, \pi^0$), 
it is given by
$ V_{ud} [H_d(x) - H_u(x)](c_+ - c_-)/\sqrt{2} $ 
by exchanging $u$ and $d$ by assuming the isospin symmetry
in the GPDs.
Therefore, the quark GPD should be written as
\begin{align}
& F^{q}(x,\xi,t) = 
\frac{1}{\sqrt{2}} \left [ \, F^u(x,\xi,t)-F^d(x,\xi,t) \right.
\nonumber \\
   &  \ \hspace{2.2cm}
    + \left.  F^u(-x,\xi,t)-F^d(-x,\xi,t)  \, \right ] 
\nonumber \\
& \ \ \ \ \ \ \ \ \ 
\text{for the $\pi^0$ production
$\nu n \to \ell^- p \, \pi^0$}. 
\label{eqn:pi0-Fq}
\end{align}
If the isospin symmetry is valid, 
both $\pi^0$ cross sections are equal 
$\sigma(\bar\nu p \to \ell^+ n \, \pi^0)
=\sigma(\nu n \to \ell^- p \, \pi^0)$.

In Eq.\,(\ref{eqn:quark-gluon-GPD-integral}),
the integral over $x$ is given by using 
the relation with the principal integral and the delta function, 
$1/(x-\xi +i\epsilon) = {\cal P} /(x-\xi)-i \pi \delta (x-\xi)$,
as 
\begin{align}
\! \! \! 
    \int_{-1}^1 \! \! dx \frac{F (x,\xi,t)}{x-\xi +i\epsilon} 
    = {\cal P} \! \! \int_{-1}^1 \! \! dx \frac{F (x,\xi,t)}{x-\xi} 
      - i \pi F (\xi,\xi,t), 
\label{eqn:x-integral}
\end{align} 
so that 
the GPDs in the whole kinematical region of $x$, 
including the ERBL region ($-\xi < x <\xi$),  
contribute to the cross section.
It could mean that the cross section is sensitive to the meson-pole 
contributions discussed in Sec.\,\ref{pion-rho-da}
because the meson-pole GPDs exist in the ERBL region.

Using these quark and gluon amplitudes with the GPDs,
we write the neutrino pion-production cross section as
\cite{psw-2017}
\begin{eqnarray}
\label{cs}
\frac{d^4\sigma(\nu N\to \ell N^\prime \pi)}{dy\, dQ^2\, dt}
 = 2 \pi \bar\Gamma \varepsilon\sigma_{L} .
\end{eqnarray}
The factor $\varepsilon$ indicates the fraction of 
the cross section for the longitudinal photon polarization,
and it is defined by
$\varepsilon =1/[1-2\vec q^{\, 2} \tan^2 (\theta/2)/q^2]$,
which is approximated as $\varepsilon \simeq (1-y)/(1-y+y^2/2)$,
with the scattering angle $\theta$ for the lepton \cite{Haltzen-1984}.
The $\bar \Gamma$ is the kinematical factor given by
\begin{align}
\! 
\bar \Gamma = \frac{G_F^2 Q^2}
{32 (2 \pi)^4 y \sqrt{ 1+4x_B^2 m_N^2/Q^2} (s-m_N^2)^2 (1-\varepsilon)}  ,
\label{eqn:gamma-bar}
\end{align}
where $G_F$ is the Fermi coupling constant.
The longitudinal cross section 
$\sigma_{L} = \epsilon_L^{*\mu} \epsilon_L^\nu W_{\mu\nu}$ 
for $W (\epsilon_L) N \to \pi N'$ is expressed as
\begin{align}
& 
\sigma_{L} =    \frac{1} {Q^2}\bigg \{ \left [\, 
      |C_q{\mathcal{H}}^q + C_g{\mathcal{H}}^g|^2 
      + |C_q\tilde{\mathcal{H}}^{q}|^2 \right ] (1-\xi^2) 
\nonumber \\[-0.20cm]
& \hspace{1.6cm} 
      +\frac{\xi^4}{1-\xi^2} \left [\,   |C_q {\mathcal{E}}^q
      +C_g {\mathcal{E}}^g |^2 
      + |C_q\tilde{\mathcal{E}}^q |^2 \right ]  
\nonumber  \\[-0.00cm]
& \hspace{1.6cm}
      -2 \xi^2 \, \text{Re} \left [ (C_q{\mathcal{H}}^q
      + C_g{\mathcal{H}}^g) ( C_q {\mathcal{E}}^q
      + C_g {\mathcal{E}}^g) ^*  \right ]
\nonumber  \\[-0.00cm]
& \hspace{1.6cm}
      -2 \xi^2 \, \text{Re} \left[
       ( C_q\tilde{\mathcal{H}}^q ) 
       ( C_q \tilde{\mathcal{E}}^q)^* \right ] \bigg \} .
\label{eqn:sigma-l-cross}
\end{align}
Although both ${\mathcal{H}}^q$ and ${\mathcal{H}}^g$ 
(${\mathcal{E}}^q$ and ${\mathcal{E}}^g$) 
exist in this equation, only the quark terms 
$\tilde{\mathcal{H}}^q$ and $\tilde{\mathcal{E}}^q$ exist 
without the gluon contributions.
As explained in Ref.\,\cite{psw-2017}, the gluon axial amplitude vanishes
by summing the considered gluon processes, so that there is no
$\tilde{\mathcal{H}}^g$ and $\tilde{\mathcal{E}}^g$ in the cross section.
The functions ${\mathcal{H}}^q$, ${\mathcal{H}}^g$, $\cdots$,
which are called Compton form factors, will be determined from the cross section 
in Eq.\,(\ref{eqn:sigma-l-cross}). These Compton form factors are related
to the corresponding GPDs by the relations in 
Eq.\,(\ref{eqn:quark-gluon-GPD-integral}).
In order to obtain the GPDs from the Compton form factors,
deconvolution is necessary, and there were recent studies 
on this issue with a shadow GPD problem \cite{deconvolution-sgpd}.
The same deconvolution issue exists in the neutrino reactions.
Once the neutrino cross sections are measured in future,
one should be careful about handling this problem.
In this way, if the neutrino cross sections are measured
for the pion production, the GPDs could be determined
by a global analysis together with the charged-lepton GPD experiments.

\section{Results}
\label{results}

\subsection{Used GPDs for calculating cross sections}
\label{gpds-gk}

The GPDs are still not well determined yet from global analysis.
There are three variables $x$, $\xi$, and $t$ for describing the GPDs.
In the current charged-lepton measurements, 
the full kinematical region cannot be investigated
including the ERBL region $-\xi < x < \xi$.
Therefore, the current versions of the GPD parametrizations need 
improvements to become a similar status of the unpolarized-PDF 
parametrizations.
Nonetheless, there are constraints on some GPDs.
For understanding them intuitively, let us consider
possibly the simplest factorized form, for example,
for the GPDs $H$ at $\xi=0$ as
\cite{gpd-paramet}
\begin{align}
H^i (x,\xi=0,t)= f_i (x) \, F_i (t, x) , \ \ 
i=q,\ g,
\label{eqn:gpd-paramet1}
\end{align}
where $f_i (x)$ is a longitudinal PDF, 
$F_i (t, x)$ is a transverse form factor at $x$,
and $i$ indicates a quark or gluon.
Therefore, the $x$ and $t$ distributions are 
constrained by the longitudinal PDFs and 
the form factors, which can be 
investigated in other experiments. 
Especially, the GPDs $H$ and $\tilde H$ should be constrained
by the unpolarized and longitudinally-polarized PDFs,
respectively.

As a useful model for the GPDs, we use 
the Goloskokov-Kroll (GK) parametrization proposed 
in a series of papers in Refs.\,\cite{GK-GPDs}.
These works were summarized in Ref.\,\cite{Kroll-2013},
so that it is sometimes called the GK-2013 parametrization.
It is accommodated as the GK model in the code 
of the PARTONS project \cite{partons-2018} for the GPDs.
We use the GK parametrization with modifications
for the gluon GPD $E_g$ and 
by including the rho-pole effects in the GPDs $H_q$ and $E_q$. 
On the GPDs $E$, their parameters are not shown in Ref.\,\cite{Kroll-2013},
so that the 2009 version of the GK is taken \cite{GK-GPDs}.
We use the GPDs $\tilde E_q$ including the pion-pole terms,
which exist in the GK parametrization,
whereas the rho-pole term does not exist in the GK.
These GPD modifications from the original GK parametrization
are explained in showing the numerical results 
in the next subsection.

\begin{figure*}[t]
 \vspace{-0.00cm}
   \includegraphics[width=18.0cm]{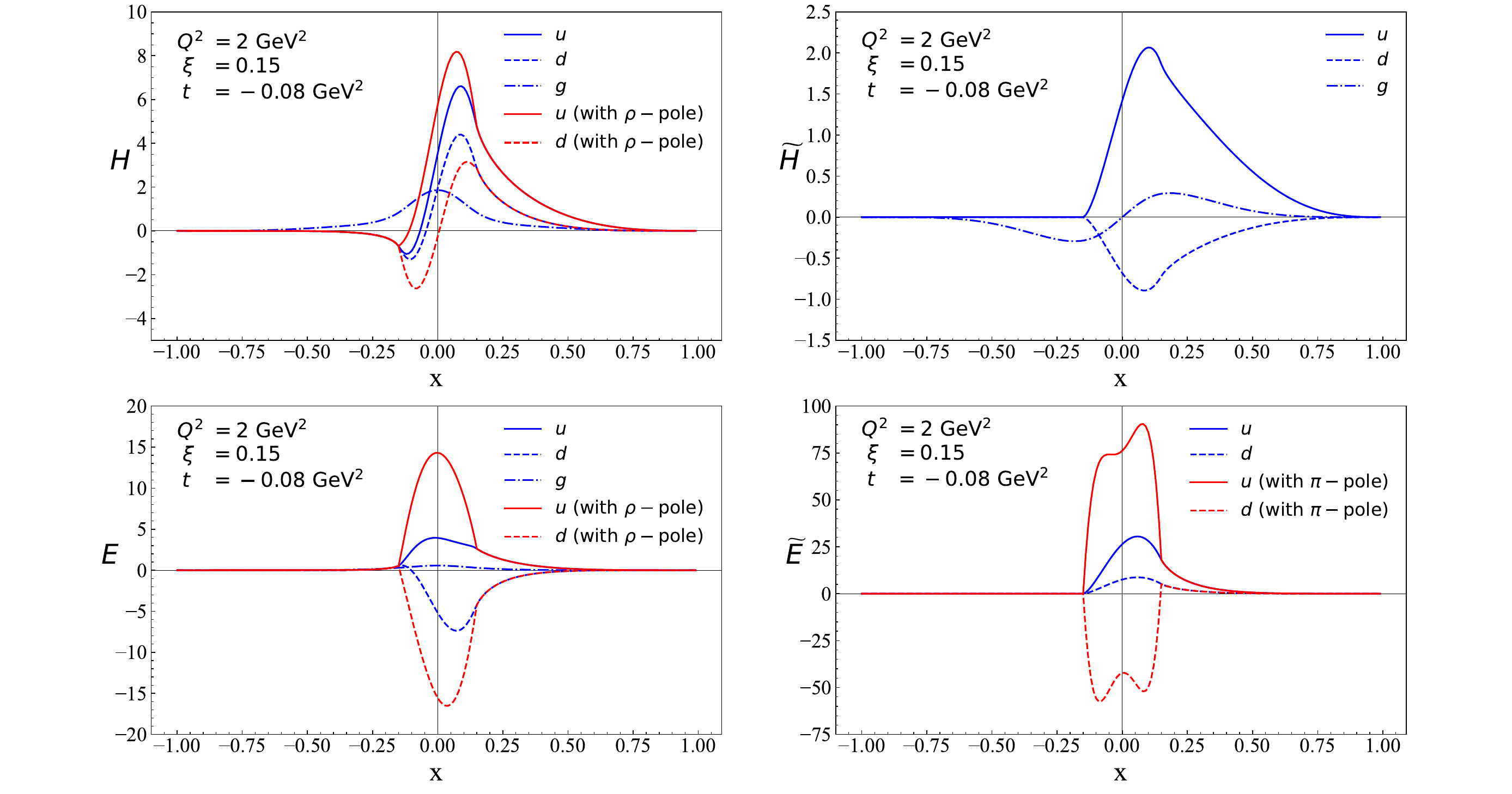}
\vspace{-0.7cm}
\caption{GPDs of the GK parametrization, 
with the additional rho-pole terms, 
are shown as the function of $x$
at $Q^2$=2 GeV$^2$, $\xi=0.15$, $t=-0.08$ GeV$^2$.
The quark GPDs $H^{u,d}$, $E^{u,d}$, and $\tilde E^{u,d}$
are shown with and without the pion- and rho-pole GPDs.
}
\label{fig:gpds-gk}
\vspace{-0.30cm}
\end{figure*}

The general form of the GK parametrization is given 
in the integral form as \cite{Kroll-2013}
\begin{align}
F^i(x,\xi,t) & = \int_{-1}^{1}\, d\rho\,\int_{-1+|\rho|}^{1-|\rho|}\, d\eta
              \,\delta(\rho+\xi\eta -x)\, f^i(\rho,\eta,t)
\nonumber \\
              & \ \ + D_i(x,t)\, \Theta(\xi^2-x^2) ,
\label{eq:int-rep}
\end{align}

where $\Theta(\xi^2-x^2)$ is the Heaviside step function
\begin{align}
\Theta(\xi^2-x^2)= 
\left \{  
    \begin{array}{l}
    1 \ \ (\xi^2 > x^2) \\
    0 \ \ (\xi^2 < x^2)
    \end{array}
\right.   .
\end{align}
In the GK-2013, the term $D_i(x,t)\, \Theta(\xi^2-x^2)$
exists only in the GPD $\tilde{E}$ because the pion-pole GPD is
included. In our numerical analysis of this work, we also include
rho-pole GPDs $H$ and $E$, so that such D terms exist in these GPDs.
The double parton distributions $f^i$ are parametrized in the following way:
\begin{align}
f^i(\rho,\eta,t) & = F^i(\rho,\xi=0,t)\, w_i(\rho,\eta) ,
\nonumber \\
F^i(\rho,\xi=0,t) & = F^i(\rho,\xi=0,t=0)\, \exp \big( \,  t  \, p_{fi}(\rho)  \, \big) ,
\nonumber \\
p_{fi}(\rho) & = -\alpha_{fi}^\prime \ln{\rho} + b_{fi} ,
\nonumber \\
w_i(\rho,\eta) & = \, \frac{\Gamma (2n_i+2)}{2^{2n_i+1}\Gamma^2 (n_i+1)} \,
\frac{[(1-|\rho |)^2-\eta^2]^{n_i}}{(1-|\rho |)^{2n_i+1}} .
\\[-0.70cm]\nonumber
\label{eq:double}
\end{align}
The parameters were determined from a global analysis.
The distributions in the negative $x$ region should be understood
by discussing valence- and sea-quark separately as
\begin{align}
F_{\rm val}^{q}(-x,\xi,t) & = 0\,, \qquad -1\leq x\leq -\xi ,
\nonumber \\
F_{\rm sea}^{q}(-x,\xi,t) &= -\epsilon_f\, F_{\rm sea}^{q}(x,\xi,t) ,
\nonumber \\
F^g(-x,\xi,t) &=\epsilon_f\, F^g(x,\xi,t) ,
\end{align}
where $\epsilon_f=1$ for the GPDs $H$ and $E$
and $\epsilon_f=-1$ for the GPDs $\tilde H$ and $\tilde E$.
The parameter values are listed in the tables of 
Refs.\,\cite{Kroll-2013,GK-GPDs}. 
In the integral of Eq.\,(\ref{eq:int-rep}), one needs to be
careful about separate treatments on valence-quark,
sea-quark, and gluon integrals as
\begin{align}
& \! \! \!
\text{valence: }
 \int_{0}^{1}\, d\rho\,\int_{-1+|\rho|}^{1-|\rho|}\, d\eta
              \,\delta(\rho+\xi\eta -x)\, f_{\rm val}^{q}(|\rho|,\eta,t)  ,
\nonumber \\
& \! \! \!
\text{sea: }
 \int_{0}^{1}\, d\rho\,\int_{-1+|\rho|}^{1-|\rho|}\, d\eta
              \,\delta(\rho+\xi\eta -x)\, f_{\rm sea}^{q}(|\rho|,\eta,t)  
\nonumber \\
& \! \! \!
\ \ \ \,
 - \int_{-1}^{0}\, d\rho\,\int_{-1+|\rho|}^{1-|\rho|}\, d\eta
              \,\delta(\rho+\xi\eta -x)\, f_{\rm sea}^{q}(|\rho|,\eta,t)  ,
\nonumber \\
& \! \! \!
\text{gluon: }
 \int_{-1}^{1}\, d\rho\,\int_{-1+|\rho|}^{1-|\rho|}\, d\eta
              \,\delta(\rho+\xi\eta -x)\, f^{g}(|\rho|,\eta,t)  .
\label{eq:integral-val-sea-gluon}
\end{align}
One can use these functions with the parameter values 
for numerical evaluations or simply use the PARTONS code
for the GK model \cite{partons-2018}.

The obtained GPDs are shown in Fig.\,\ref{fig:gpds-gk}
by taking $Q^2$=2 GeV$^2$, $\xi=0.15$, $t=-0.08$ GeV$^2$.
The The $u$-quark, $d$-quark, and gluon distributions are shown
except for $\tilde E$, where there is no gluon distribution.
Because the unpolarized antiquark distribution functions
have been determined well from global analyses,
there are strong constraints on the $x$ dependence of the GPDs
$H$ for antiquarks. Therefore, the GK GPDs contain the sea-quark
component in addition to the valence-quark and gluon GPDs. 
It is the reason why there are finite distributions
at $x<-\xi=-0.15$ in the function $H^{q}$.
On the other hand, there is little information
on sea-quark GPDs on other functions $\tilde H$, $E$, and $\tilde E$.
Therefore, the GK parametrization supplies only the valence-quark
GPDs for the quark part in these GPDs, which indicates that the quark GPDs
are zero at $x<-\xi=-0.15$ as clearly shown in Fig.\,\ref{fig:gpds-gk}.
Because of this, there are sudden changes at $x=-\xi=-0.15$ 
in the quark GPDs $\tilde H^q$, $E^q$, and $\tilde E^q$.

We also notice that the pion- and rho-pole contributions
are large in all the GPDs of $H$, $E$, and $\tilde E$
in Fig.\,\ref{fig:gpds-gk}. In particular, 
the pion-pole effects are much larger than other GPD distributions
in $\tilde E$, and the rho-pole effects are much larger 
than other distributions in $E$.
The rho-pole effects are comparable in magnitude 
to the other distributions in $H$.
These facts indicate that the understanding of the meson
contributions is important for clarifying
the GPDs in the whole kinematical region.

As for the gluon GPDs, the unpolarized gluon distribution
is relatively well determined, so that there are constrains
on the $x$ distribution part of the gluon GPD $H^g$.
However, the longitudinally-polarized gluon distribution
still has a large uncertainty and it cannot be fixed
in the current global analysis. 
Therefore, although $\tilde H^g$ distribution of Fig.\,\ref{fig:gpds-gk}
is supplied in the GK parametrization, it could have a large uncertainty.
As for the gluon GPD $E^g$, there is little information, 
so that its distribution should be considered as 
an undermined quantity at this stage although
$E^g$, which is the variant 3 of the GK-2009 paper
\cite{GK-GPDs}, is shown in Fig.\,\ref{fig:gpds-gk}.
The $E^g$ distribution is much smaller than other quark GPDs $E^{u,d}$. 
There is no $\tilde E^g$ in the GK parametrization
at this stage.

\vspace{-0.20cm}

\subsection{Cross sections for the $\pi^+$ production}
\label{pi+}

\begin{figure}[b]
 \vspace{-0.30cm}
\begin{center}
   \includegraphics[width=8.5cm]{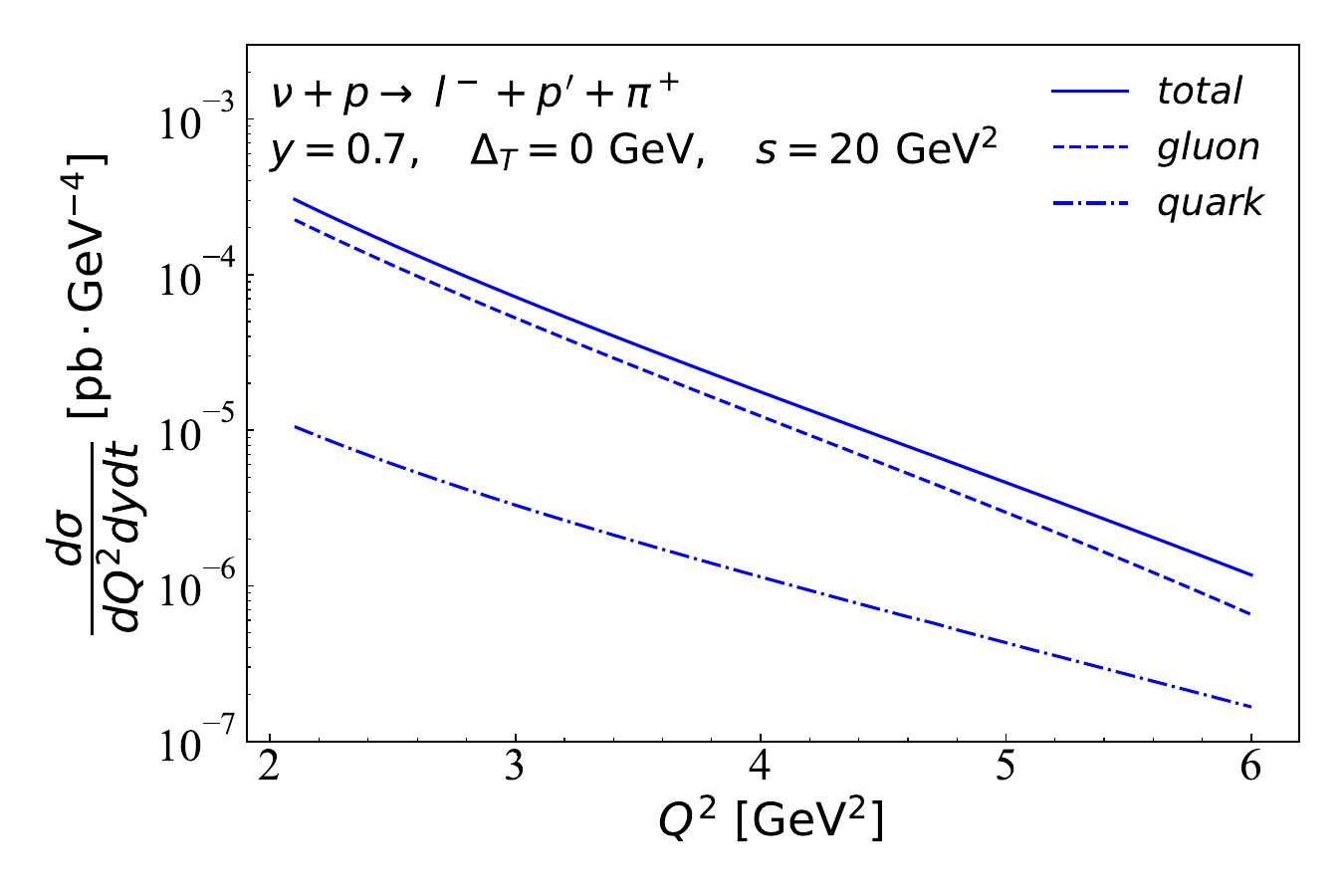}
\end{center}
\vspace{-0.9cm}
\caption{$\pi^+$-production cross sections for 
$\nu  p \to \ell^-  p' \, \pi^+$
at $y=0.7$, $\Delta_T =0$ GeV, and $s=20$ GeV$^2$. }
\label{fig:pi+-cross-sections}
\vspace{-0.00cm}
\end{figure}

\begin{figure*}[t]
 \vspace{-0.00cm}
   \includegraphics[width=18.0cm]{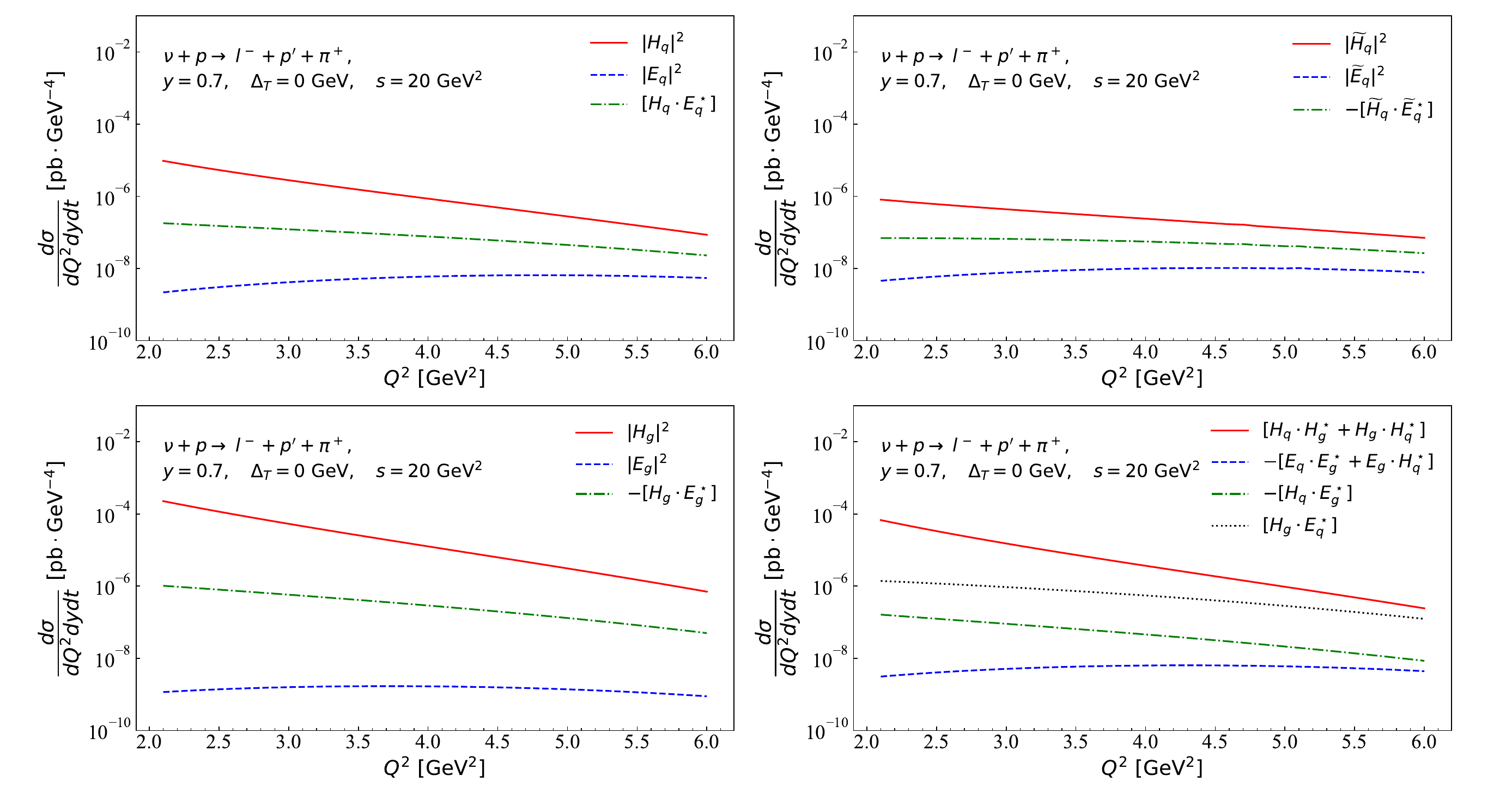}
\vspace{-0.7cm}
\caption{Each component of the $\pi^+$-production cross section
for $\nu  p \to \ell^-  p' \, \pi^+$.}
\label{fig:pi+-each-component}
\vspace{-0.30cm}
\end{figure*}

First, we show the $\pi^+$ production cross section by
using the GK GPDs of Sec.\,\ref{gpds-gk}.
The same kinematical condition of Ref.\,\cite{psw-2017} is taken
for calculating the cross section, because the condition
could be realized roughly in the Fermilab experiment with 
the 10 GeV neutrino beam. Namely, we take $s=20$ GeV$^2$, $y=0.7$,
and $\Delta_T=0$. This choice of $\Delta_T=0$ means 
that the momentum-transfer squared $t$ is expressed
by $\xi$ as follows.
From Eq.\,(\ref{eqn:xi-x}), the variable $t$ is expressed 
in terms of $\xi$ and the transverse component
 of the momentum transfer $\Delta_T$ and it becomes
$ t  = - 4 \xi^2 m_N^2 / (1-\xi^2) + \Delta_T^2/(1-\xi^2)  
     =  - 4 \xi^2 m_N^2 /(1-\xi^2)$
if $\Delta_T^2=-\vec\Delta_T^{\,2}=0$.
Later in this subsection, the $s$ dependence of the cross section 
is shown by taking also $s=30$ and 100 GeV$^2$.

The $\nu + p \to \ell^- + p' + \pi^+$ cross section is shown
in Fig.\,\ref{fig:pi+-cross-sections} 
as the function of $Q^2$ from 2 GeV$^2$ to 6 GeV$^2$.
Three curves are shown. The solid curve indicates the cross
section with the whole quark and gluon GPD contributions,
the dashed does the gluon GPD contribution by terminating
the quark GPDs ($C_q=0$), and
the dot-dashed does the quark GPD contribution by terminating
the gluon GPDs ($C_g=0$).
We notice that the whole cross section is dominated by
the gluon GPD terms.
This gluon dominance is roughly understood as follows.
Let us consider the kinematical factors 
in the quark and gluon contributions. 
In the cross section of Eq.\,(\ref{eqn:sigma-l-cross}),
there are coefficients $C_q$ and $C_g$ in front of
the quark and gluon Compton form factors, which
contain the factors $2 f_\pi$ and $8 f_\pi /\xi$
and the integrals of pion distribution amplitude
in Eq.\,(\ref{eqn:quark-gluon-GPD-integral}).
The ratio of these factors are
\begin{align}
\frac{\xi C_q}{8 C_g}
     =0.05 - 0.19 
 \ \ \ \text{for $y=0.7$} 
\label{eqn:q-g-ratio}
\end{align}
from $Q^2=$2 GeV$^2$ to 6 GeV$^2$.
In Fig.\,\ref{fig:pi+-cross-sections}, the quark GPD contributions
to the cross sections are one order of magnitude smaller 
than the gluon GPDs ones, which roughly corresponds this ratio.
However, the functional forms are different between the quark 
and gluon GPDs, which leads to another factor in estimating 
the precise quark-gluon contribution ratio.

In order to find which GPDs will be determined from
the neutrino reactions, we show each component contribution
to the $\pi^+$ cross section in Fig.\,\ref{fig:pi+-each-component}.
Namely, the cross sections are shown from 
\begin{itemize}
\vspace{-0.10cm}
\setlength{\itemsep}{-0.05cm} 
\item[(1)] quark GPD terms of $H^q$ and $E^q$:\\[+0.1cm]
$|\mathcal{H}^q|^2$, 
$|\mathcal{E}^q|^2$, 
$\mathcal{H}^q \mathcal{E}^{q*}$,
\item[(2)] quark GPD terms of $\tilde H^q$ and $\tilde E^q$:\\[+0.1cm]
$|\tilde{\mathcal{H}}^q|^2$,
$|\tilde{\mathcal{E}}^q|^2$,
$\tilde{\mathcal{H}}^q \tilde{\mathcal{E}}^{q*}$,
\item[(3)] gluon GPD terms of $H^g$ and $E^g$:\\[+0.1cm]
$|\mathcal{H}^g|^2$, 
$|\mathcal{E}^g|^2$, 
$\mathcal{H}^g \mathcal{E}^{g *}$,
\item[(4)] quark-gluon interference terms:\\[+0.1cm]
$\mathcal{H}^q \mathcal{H}^{g*}+ \mathcal{H}^g \mathcal{H}^{q*}$, 
$\mathcal{E}^q \mathcal{E}^{g*}+ \mathcal{E}^g \mathcal{E}^{q*}$, \\[+0.1cm]
$\mathcal{H}^q \mathcal{E}^{g*}$,
$\mathcal{H}^g \mathcal{E}^{q*}$,
\vspace{-0.15cm}
\end{itemize}

for the $\pi^+$-production cross section.

First, the largest contribution comes from the gluon GPD $H^g$
as shown in the curve of $|\mathcal{H}^g|^2$, whereas
the $E_g$ term $|\mathcal{E}^g|^2$ is much smaller 
than the leading $|\mathcal{H}^g|^2$.
Their interference $\mathcal{H}^g \mathcal{E}^{g *}$
is typically one order of magnitude smaller
than $|\mathcal{H}_g|^2$. 
It means that the $\pi^+$ cross section is mainly sensitive
to the gluon GPD $H^g$.
As shown in Eq.\,(\ref{eqn:sigma-l-cross}),
there is no term from the polarized GPDs
$\tilde H^g$ and $\tilde E^g$.
These results depend, of course, on the employed GPDs.
In particular, the gluon GPD $E^g$ is weakly constrained
in the current parametrization and it should have 
larger uncertainty even in the overall magnitude.

Second, the pure quark-term contributions are one or two orders of 
magnitude smaller than the gluon ones. For example,
the $|\mathcal{H}^q|^2$ cross section is one order 
of magnitude smaller than the leading $|\mathcal{H}^g|^2$ one.
In the same way with the gluon case, the $|\mathcal{E}^q|^2$
is much smaller than $|\mathcal{H}^q|^2$, so the leading
quark contribution comes from $H^q$.
The polarized quark GPDs' contributions are about 
one order of magnitude smaller than the corresponding
unpolarized quark GPDs.
For example, the $|\tilde{\mathcal{H}}^q|^2$ cross section 
is one order of magnitude smaller than the leading $|\mathcal{H}^q|^2$ one.
The $|\tilde{\mathcal{E}}^q|^2$ cross section is further smaller
than the $|\tilde{\mathcal{H}}^q|^2$ one.

Third, the quark-gluon interference contributions are 
between the pure quark and gluon cross sections in magnitude.
Among the terms, the GPD term
$\mathcal{H}^q \mathcal{H}^{g*}+ \mathcal{H}^g \mathcal{H}^{q*}$
is the largest one, the next one is $\mathcal{H}^g \mathcal{E}^{q*}$,
and then $\mathcal{H}^q \mathcal{E}^{g*}$. 
The $\mathcal{E}^q \mathcal{E}^{g*}+ \mathcal{E}^g \mathcal{E}^{q*}$ 
term is the smallest in the interference.
From these discussions, we roughly list the GPDs from the largest as
\begin{align}
(1) \ H^g \ \ \ (2) \ H^q \ \ \ 
(3) \ \tilde H^q \ \ \ 
(4) E^q, \ \tilde E^q, \ E^g ,
\label{eqn:order-contributions}
\end{align}
for the $\pi^+$ production. In the $\pi^0$ production, 
the gluon GPDs do not contribute as the leading one
as discussed later in this section.

\begin{figure}[t]
 \vspace{+0.17cm}
\begin{center}
   \includegraphics[width=8.5cm]{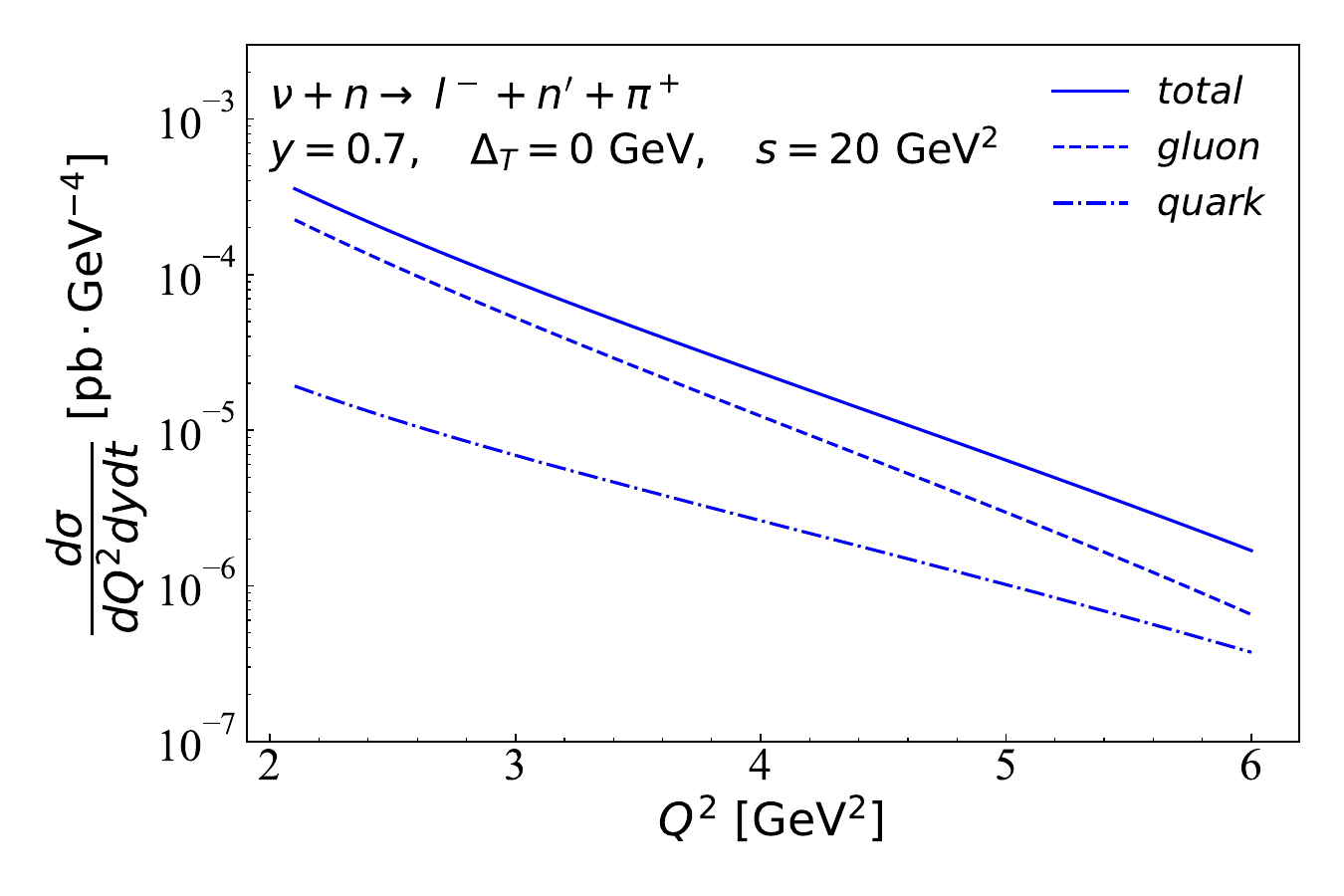}
\end{center}
\vspace{-0.9cm}
\caption{
$\pi^+$-production cross sections for 
$\nu  n \to \ell^-  n' \, \pi^+$
at $y=0.7$, $\Delta_T =0$ GeV, and $s=20$ GeV$^2$.
}
\label{fig:pi+-cross-sections-n}
\vspace{-0.30cm}
\end{figure}

As shown in Fig.\,\ref{fig:pi+-cross-sections} and also later 
in Fig.\,\ref{fig:pi+-cross-sections-n}, the fraction of 
the quark-GPD contribution increases at large $Q^2$.
This tendency is understood from Fig.\,\ref{fig:pi+-each-component}, 
where the gluon-$H$ term drops more rapidly as increasing $Q^2$ 
than quark-$H$ term does mainly due to the kinematical
factor $\xi$ given in Eq.\,(\ref{eqn:q-g-ratio}).
The ratio of Eq.\,(\ref{eqn:q-g-ratio}) increases from
0.05 at $Q^2=$2 GeV$^2$ to 0.19 at $Q^2=$6 GeV$^2$.
In addition, there are effects from the difference 
between the functional forms of $H_g$ and $H_q$. 
We also notice that the $Q^2$ dependencies of other terms 
are different between gluon and quark contributions.

Another interesting point of the GPD determination is on the role
of the GPDs in the ERBL region. After Eq.\,(\ref{eqn:x-integral}),
we mentioned that the cross section is sensitive to
the GPDs of the ERBL region. However, such effects need to be clarified
numerically for proposing neutrino GPD experiments. 
Although the ERBL is an important kinematical region,
its GPDs have not been determined  experimentally.
An ERBL-GPD investigation is under consideration, for example, 
at J-PARC by the $N+N \to \pi + N +B$ reactions
\cite{kss-2009,J-PARC-GPDs,transition-GPDs}.
We need more studies whether the GPDs in the ERBL region 
could affect the neutrino-induced pion-production cross sections
and also deeply virtual meson production (DVMP) ones 
of the charged-lepton reactions.

\begin{figure}[b]
 \vspace{-0.30cm}
\begin{center}
   \includegraphics[width=8.5cm]{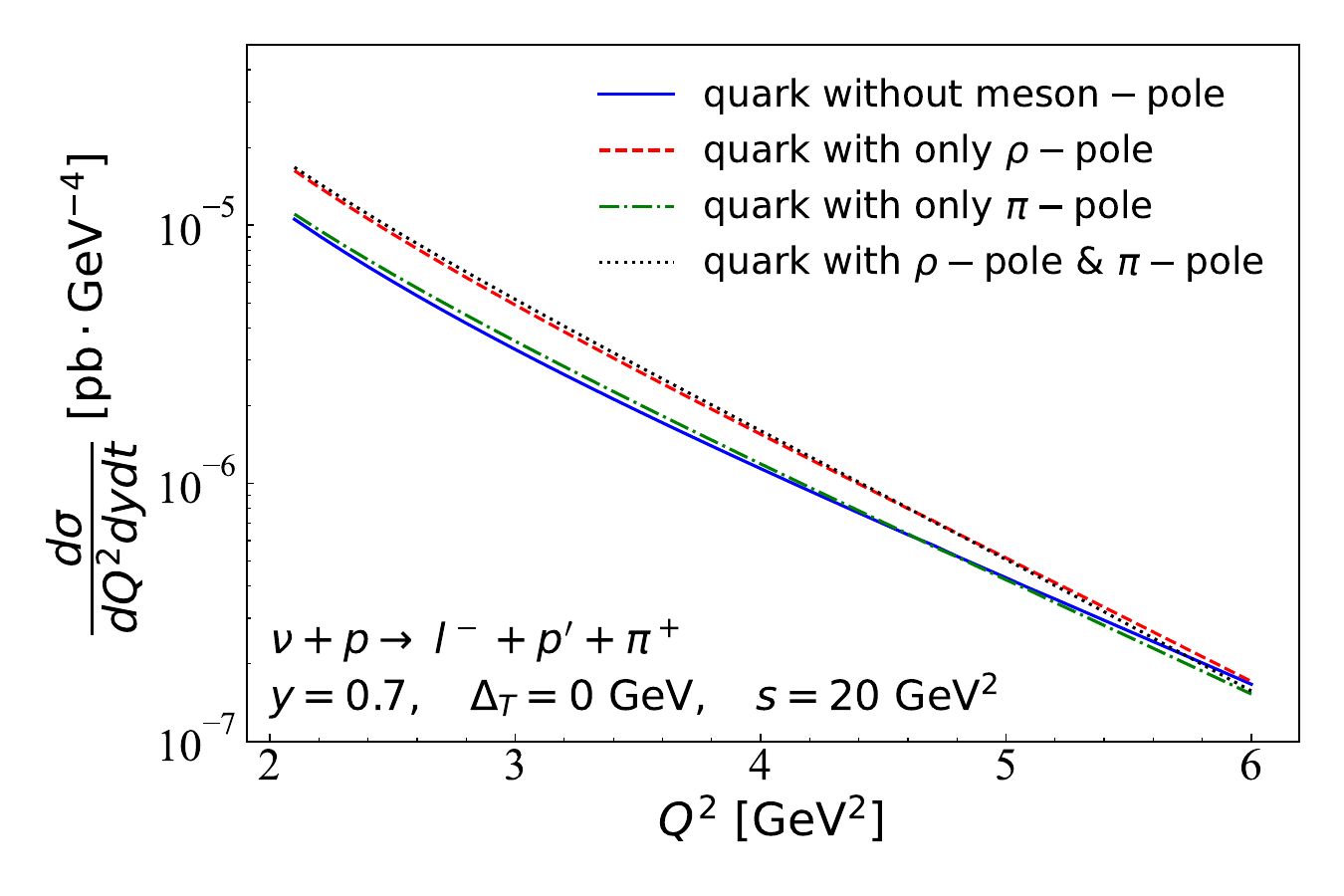}
\end{center}
\vspace{-0.9cm}
\caption{Pion- and rho-pole effects on the $\pi^+$ cross section
for $\nu p \to \ell^- p' \, \pi^+$ by terminating the gluon GPDs.}
\label{fig:pion-rho-poles}
\vspace{-0.00cm}
\end{figure}

Using the pion- and rho-pole GPDs 
\begin{align}
\tilde E_\pi, \ \ 
       H_\rho, \ \ 
       E_\rho, 
\nonumber
\end{align}
in 
Eqs.\,(\ref{eqn:pion-pole}) and (\ref{eqn:rho-pole}),
we calculated their effects on the $\pi^+$ cross section.
The results are shown in Fig.\,\ref{fig:pion-rho-poles}
by terminating the gluon GPDs in order to clarify 
these pole-term contributions.
Four curves are shown by taking the following
four types of the GPDs:
\begin{itemize}
\vspace{-0.10cm}
\setlength{\itemsep}{-0.05cm} 
\item[(1)] GPDs with the pion and rho poles,
\item[(2)] GPDs with only the pion pole,
\item[(3)] GPDs with only the rho pole,
\item[(4)] GPDs without pion and rho poles .
\vspace{-0.10cm}
\end{itemize}

The pion-pole GPDs ($\tilde E^{u,d}_\pi$)
increase the cross section slightly
in Fig.\,\ref{fig:pion-rho-poles}.
On the other hand, the rho-pole GPD 
($H^{u,d}_\rho$, $E^{u,d}_\rho$) effects are large 
and they increase the cross section significantly. 
It simply comes from the fact that the $H^q$ effects
are much larger than the $\tilde E^q$ ones
in the cross section as shown in Eq.\,(\ref{eqn:order-contributions}).
In fact, the quark GPDs $H^q$ are significantly modified
if the rho-pole GPDs are added as shown
in Fig.\,\ref{fig:gpds-gk}.
The rho- and pion-pole effects are much larger in $E^q$
and $\tilde E^q$, respectively, in Fig.\,\ref{fig:gpds-gk}; 
however, their modifications do not appear clearly in the cross 
section simply because the fractions of $E^q$
and $\tilde E^q$ are small in the cross section.
In any case, because the pion production cross section is sensitive
to the GPDs in the ERBL region, the detailed GPD studies 
in a wide kinematical region could becomes possible 
by investigating the neutrino reactions in addition 
to the charged-lepton reactions.
In general, the pure quark-GPD contributions are about 
one order of magnitude smaller than the pure-gluon GPD ones. 
However, the quark-gluon interference contributions
are not so small as shown in the lower-right figure of 
Fig.\,\ref{fig:pi+-each-component}, so that 
the meson-pole effects may appear in measured $\pi^+$-production 
cross sections. These effects should be
more pronounced in the $\pi^0$ production in Sec.\,\ref{pi0} 
because the gluon contribution does not exist.

The pion-pole term is included in the GK parametrization 
\cite{Kroll-2013,GK-meson-pole},
and the GPD constraints are satisfied with this term. For example,
the first moment of the pion-pole term becomes the pseudoscalar form factor
as shown in Eq.\,(\ref{eqn:pi0-Fq}) of Ref.\,\cite{Vanderhaeghen-1999}.
The pion-pole term could contribute to the DVCS; 
however, it is very small due to the small
$\pi^0 \gamma \gamma$ coupling \cite{Goloskokov-2024}.
We need to make a new global analysis in principle
if the rho-pole term is included in the GPDs for the accurate
determination of the GPDs because the rho-pole term is not included
in the GK parametrization. We consider it as a future project.
At this stage, we simply added the rho-pole term to the existing
GPD parametrization in order to find its effects on the neutrino
cross section. 
One may note that there is not much freedom in
the meson-pole terms of Eqs.\,(\ref{eqn:pion-pole}) and (\ref{eqn:rho-pole}), 
although the t-dependent form factor could deviate from the dipole forms 
at large $|t|$. Therefore, we think that such pole terms exist 
in the close forms of Eqs.\,(\ref{eqn:pion-pole}) and (\ref{eqn:rho-pole}).
However, such additional rho-pole GPD terms contribute also the DVMP
cross sections in charged-lepton scattering, so that the whole GPDs
should be readjusted by a new global analysis. 

In the unpolarized collinear PDFs, the pion- and rho-cloud contributions 
could be calculated, for example, in discussing a possible model 
to explain $\bar u -\bar d$ and the Gottfried-sum-rule violation
\cite{ubar-dbar}. However, the global-analysis PDFs have certain 
$x$-dependent function forms with parameters, and the meson-cloud effects 
are effectively included in the PDFs.
In the same way, the rho-pole GPDs should be included within the GPD functions
in future global analyses. Currently, the GPD global analysis is
at the transition stage in the sense the pion-pole GPDs are 
separately defined in the GK parametrization. 
In our work, the rho-pole GPDs are separately treated in the similar way.
In future, the whole GPDs should be determined without replying on 
a specific model.

\begin{figure}[b]
 \vspace{-0.30cm}
\begin{center}
   \includegraphics[width=8.5cm]{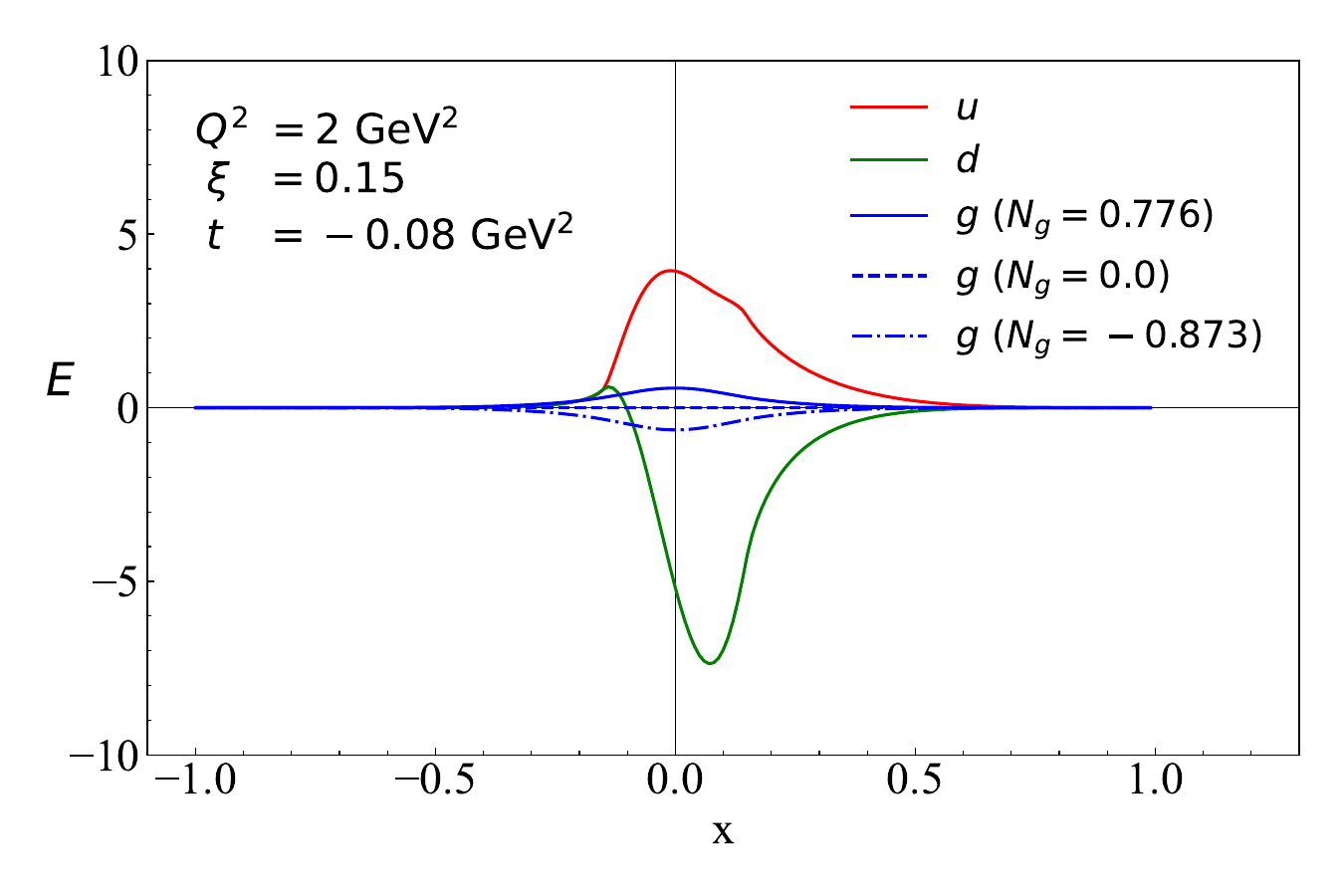}
\end{center}
\vspace{-0.9cm}
\caption{The GPDs $E^{u,d,g}$ of the GK parametrization
by taking the variants 1, 2, and 3 for $E_g$
at $Q^2 = 2$ GeV$^2$, $\xi=0.15$, and $t=-0.08$ GeV$^2$.
Here, the rho-pole terms are not included in the quark GPDs.}
\label{fig:Eg-variants}
\vspace{-0.00cm}
\end{figure}

Next, we show the effect of the gluon GPD $E^g$ on the cross section
because the $\pi^+$ production is sensitive to the gluon GPDs.
However, the gluon GPD $E^g$ could be small in comparison with
the quark ones $E^u$ and $E^d$ as already shown in Fig.\,\ref{fig:gpds-gk}.
In the GK parametrization of 2009 \cite{GK-GPDs}, the six types,
variants 1, 2, $\cdots$, 6, are prepared as different functional forms 
of $E_g$. Among them, we take the variants 1, 2, and 3 as three extreme cases
for $E_g$, and they are shown in Fig.\,\ref{fig:Eg-variants}
in comparison with the quark GPDs $E^{u,d}$, where 
the rho-pole GPDs of Fig.\,\ref{fig:gpds-gk} are not included.
In the PARTONS code \cite{partons-2018}, the GPD $E^g$
of the variant 3 is given.

Although the gluon GPD $H^g$ is as large as the quark GPDs $H^q$
as shown in Fig.\,\ref{fig:gpds-gk}, the situation is very different
for the GPDs $E^{q,g}$. The gluon GPD $E^g$ is expected to be 
much smaller than the quark GPDs $E^q$ \cite{GK-GPDs}.
The gluon GPD $E^g$ is parametrized as 
$ E^g(x,\xi=0,t=0) \equiv e_g (x) = N_g x^{-\alpha_g} (1-x)^{\beta_g} $
in the forward limit $\xi=t=0$, and the same functional forms 
are used also for the valence-quark and antiquark GPDs
[$e_{q_v}(x)$ and $e_{\bar q}(x)$].
The overall magnitudes of the valence-quark functions 
$e_{u_v}(x)$ and $e_{d_v}(x)$ are constrained by the Pauli form factors,
namely the anomalous magnetic moments ($\kappa_{p,n}$), 
of the proton and neutron as $\kappa_{u,d} = \int_0^1 dx \, e_{u_v,d_v} (x)$,
where $\kappa_u = 2\kappa_p+\kappa_n$, $\kappa_d = \kappa_p+2\kappa_n$
\cite{Diehl-2005}.
From the conservations of the momentum and total angular momentum,
we have the sum rule \cite{Diehl-Kugler-2007,GK-GPDs}
\begin{align}
\int_0^1 dx x e_g (x) = - \sum_q \int_0^1 dx x [e_{q_v}(x)+2 e_{\bar q}(x)] ,
\label{eqn:Eg-constraint}
\end{align}
for constraining the gluon part.
However, the analysis of the Pauli form factors indicated
that the $e_{u_v}(x)$ and $e_{d_v}(x)$ distributions almost cancel
with each other \cite{Diehl-2005}, which could lead to a small distribution 
of $e_g (x)$ in comparison with the overall magnitudes of 
$e_{u_v}(x)$ and $e_{d_v}(x)$ although the magnitude of
the gluon function $e_g (x)$ still depends on the antiquark functions 
$e_{\bar q} (x)$.
This situation is similar in the updated analysis of 2020
\cite{Kroll-2020}, and $e_g (x)$ of the 2020 version
is within the range between the variants 2 and 3 in Fig.\,\ref{fig:Eg-variants}.

In order to show the variations of the cross section from
the gluon GPD $E^g$, we take three GPD models, variants
1, 2, and 3, in the GK parametrization of 2009 \cite{GK-GPDs}:
\begin{itemize}
\vspace{-0.10cm}
\setlength{\itemsep}{-0.05cm} 
\setlength{\leftskip}{2.20cm}
\item[Variant 1:]\  $e_g (x) = 0$,
\item[Variant 2:]\ $e_g (x) = -0.873 \, x^{-1.1} (1-x)^6$,
\item[Variant 3:]\ $e_g (x) = \ \  0.776 \, x^{-1.1} (1-x)^6$.
\vspace{-0.10cm}
\end{itemize}
The Diehl-Kugler model-1 \cite{Diehl-Kugler-2007} and 
the Kroll-2020 model \cite{Kroll-2020} for $E^g$
are smaller in magnitude than the $E^g$ of 
the GK-variants 2 and 3.
In Fig.\,\ref{fig:Eg-variants}, these GK GPDs $E^g$
are shown with the quark GPDs 
at $Q^2 = 2$ GeV$^2$, $\xi=0.15$, and $t=-0.08$ GeV$^2$.
Here, the rho-pole terms are not included in showing the quark GPDs,
so that they are the original GK quark GPDs.
We find that the $E^g$ variations are small in comparison
with the quark GPDs.

\begin{figure}[b]
 \vspace{-0.30cm}
\begin{center}
   \includegraphics[width=8.5cm]{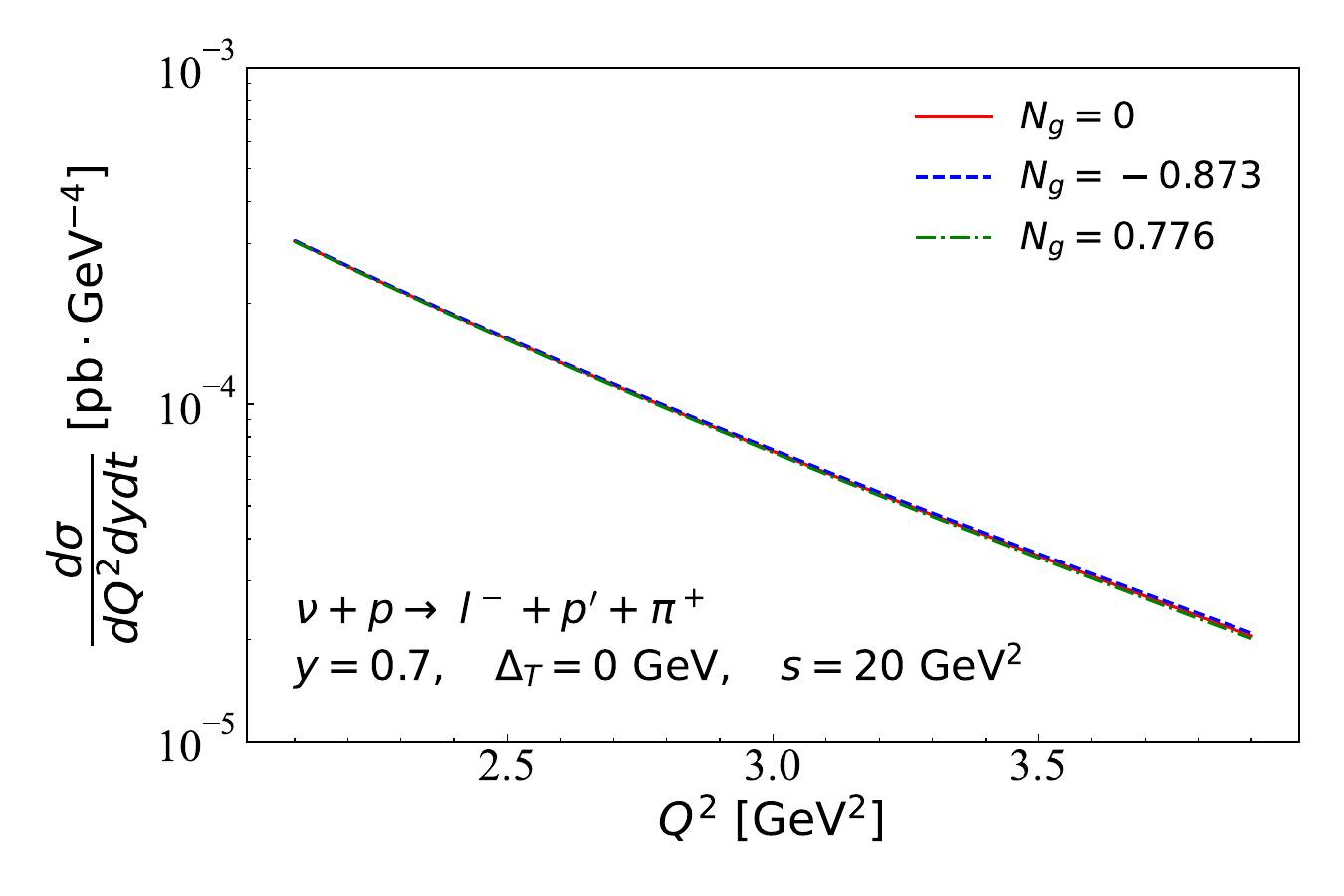}
\end{center}
\vspace{-0.9cm}
\caption{$E^g$ effects on the $\pi^+$ production cross section
for $\nu p \to \ell^- p' \, \pi^+$
by taking three GK $E^g$ models of the variants 1, 2, and 3.}
\label{fig:cross-pi+-Eg-variations}
\vspace{-0.00cm}
\end{figure}

Using these GK GPDs $E^g$, we calculated the $\pi^+$ cross sections
and they are shown in Fig.\,\ref{fig:cross-pi+-Eg-variations}.
The effects of the $E^g$ is very small in the $\pi^+$ cross
section, simply because the magnitude of $E^g$ is much smaller
than other GPDs $H^{g,q}$ and even than $E^q$.
In the $\pi^+$ cross section, the quark term
$E^d (x,\xi,t) - E^u (-x,\xi,t)$ 
contributes as shown in Eq.\,(\ref{eqn:pi+-production-quark-GPD}).
Therefore, $E^d$ and $-E^u$ additively contribute to the cross section,
which makes the $E^g$ fraction even smaller.
Therefore, it is difficult to find the gluon $E^g$
from the pion-production process in neutrino scattering.
In the pion-production measurements with unpolarized targets,
the GPD $\tilde H$, $\tilde E$ and $E$ contributions are very small.
For determining these GPDs, we should rely, for example, 
on vector-meson productions with polarized targets 
\cite{GK-GPDs,GK-meson-pole}.

\begin{figure}[t]
 \vspace{-0.00cm}
\begin{center}
   \includegraphics[width=8.5cm]{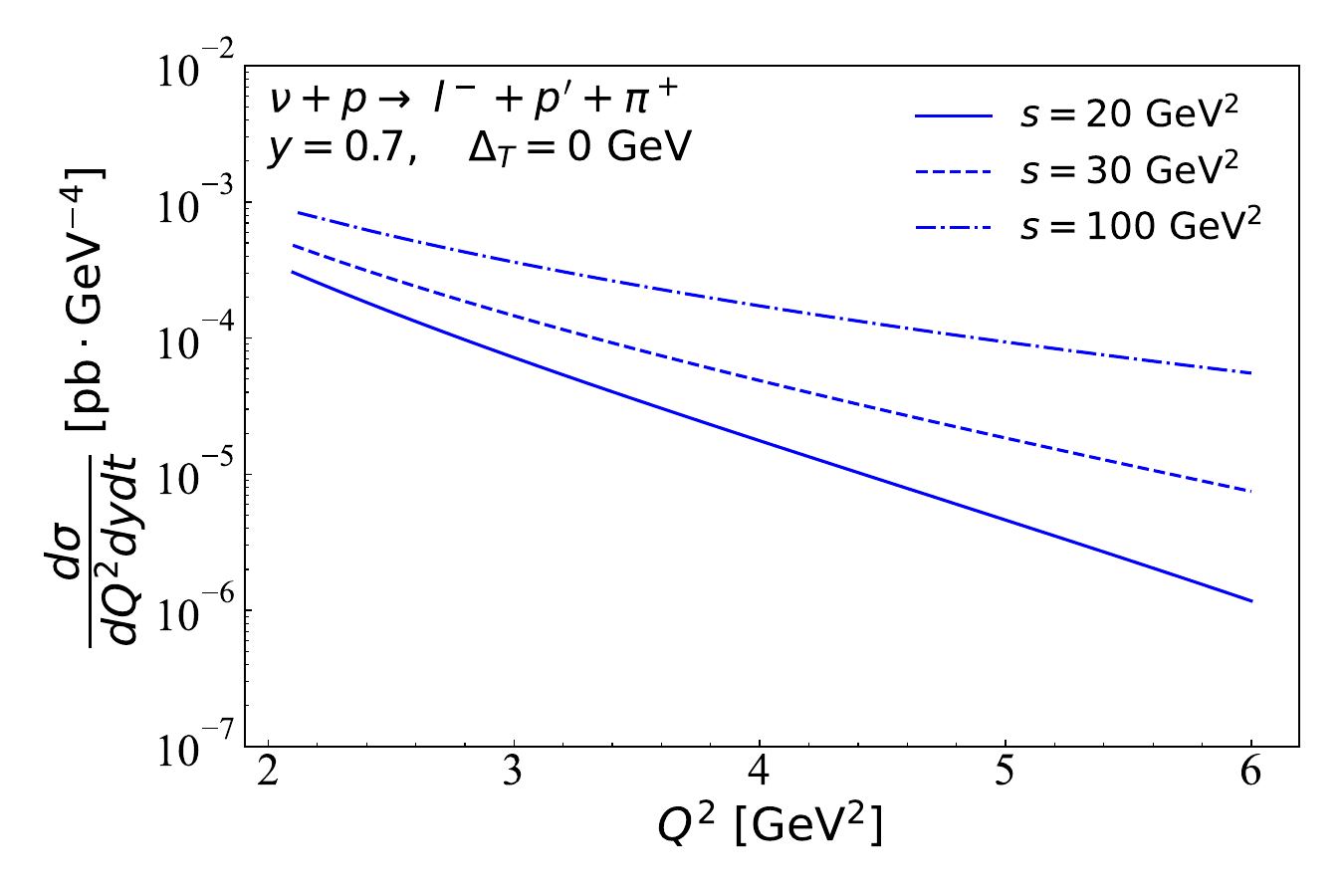}
\end{center}
\vspace{-0.9cm}
\caption{
Center-of-mass-energy squared dependence of 
the $\pi^+$-production cross section
for $\nu p \to \ell^- p' \, \pi^+$
by taking $s=20$, 30, and 100 GeV$^2$.
}
\label{fig:pi+crross-energy-dep}
\vspace{-0.30cm}
\end{figure}

In Fig.\,\ref{fig:pi+crross-energy-dep}, the $\pi^+$-production 
cross sections are shown for $\nu p \to \ell^- p' \, \pi^+$
by varying the center-of-mass-energy squared as
$s=20$, 30, and 100 GeV$^2$. 
The cross section is very sensitive to the neutrino beam energy,
and the reaction $\nu p \to \ell^- p' \, \pi^+$ cross section 
increases with the increasing energy.
The flux factor $\bar\Gamma$ of Eq.\,(\ref{eqn:gamma-bar})
decreases with the increasing $s$. However,
the longitudinal cross section $\sigma_L$ of Eq.\,(\ref{eqn:sigma-l-cross}) 
increases more rapidly with $s$ by the following reason.
The increase of $s$ means the decrease of $x$ and $\xi$, 
so that the smaller $x$ and $\xi$ regions of the GPDs contribute
to the integrals in Eq.\,(\ref{eqn:quark-gluon-GPD-integral}).
In the reaction $\nu p \to \ell^- p' \, \pi^+$,
the gluon GPDs dominate the cross section.
As $x$ becomes smaller, the gluon GPDs increase rapidly, which
leads the larger cross sections with increasing $s$.

\begin{figure}[b]
 \vspace{-0.20cm}
\begin{center}
   \includegraphics[width=7.5cm]{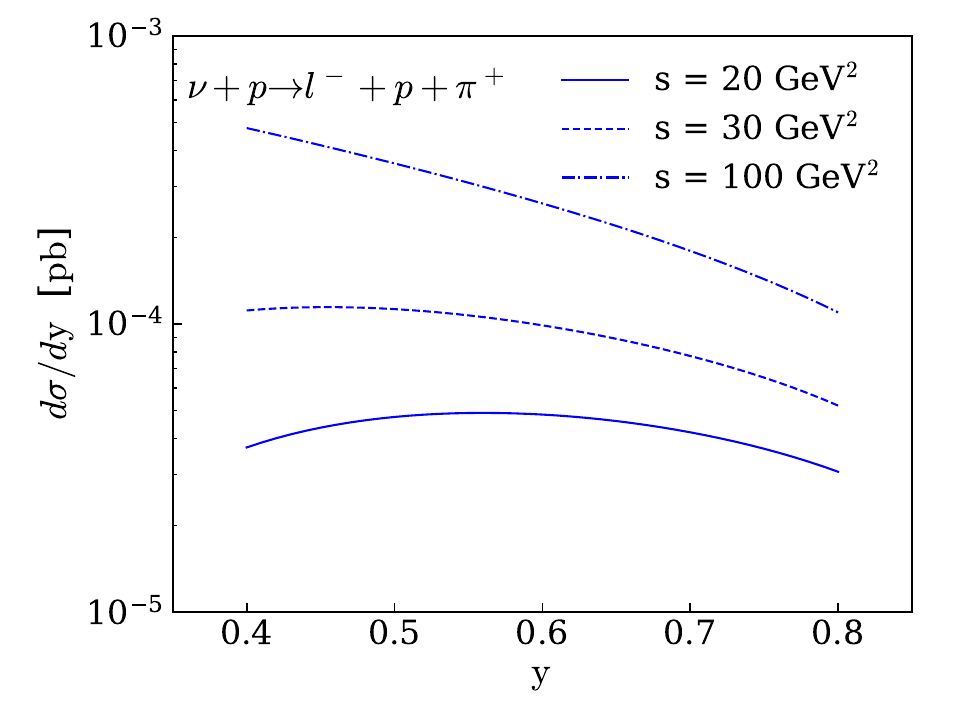}
\end{center}
\vspace{-0.7cm}
\caption{Single differential cross section $d\sigma /dy$
is shown for $\nu p \to \ell^- p \, \pi^+$
by taking $s=20$, 30, and 100 GeV$^2$.}
\label{fig:dsdy-pi+}
\vspace{-0.00cm}
\end{figure}

The single differential cross section $d\sigma /dy$
is shown for the process $\nu p \to \ell^- p \, \pi^+$
by integrating the cross section over $Q^2$ and $t$
at $s=20$, 30, and 100 GeV$^2$ in Fig.\,\ref{fig:dsdy-pi+}.
The cross section is about 
$4\times 10^{-5}$--$5 \times 10^{-4}$\,pb
at $y=0.4$--$0.5$, and it becomes smaller with increasing $y$.
The minimum value of $Q^2$ ($Q^2_{\rm min}$) for the integral 
is taken as $Q^2_{\rm min}=2$ GeV$^2$.
The cross section varies depending on $Q^2_{\rm min}$, for example,
$d\sigma /dy =3.7 \times 10^{-5}$, $4.0 \times 10^{-6}$, 
$4.2 \times 10^{-7}$\,pb
for $Q^2_{\rm min}=$2, 3, 4 GeV$^2$, 
respectively, at $y=0.4$ and $s=20$ GeV$^2$.
The $y$ dependence of the cross section is contained mainly in
the longitudinal photon polarization $\varepsilon$
above Eq.\,(\ref{eqn:gamma-bar}) as $2(1-y)$ at large $y$
and in the $\bar\Gamma$ of Eq.\,(\ref{eqn:gamma-bar}) as $1/y$,
so that the cross sections decrease with increasing $y$ in general.
The variable $y$ is related to $x$ and $Q^2$ as
$y=Q^2/[x(s-m_N^2)]$, so that $x$ dependence of the GPDs
also affect.

The cross section for $\pi^+$ production from the neutron
($\nu  n \to \ell^-  n' \, \pi^+$)
is also calculated in the same way by exchanging the GPDs
$F^u$ and $F^d$ from the proton-target case
due to the isospin symmetry.
Its cross section is shown in
Fig.\,\ref{fig:pi+-cross-sections-n} by showing
the quark and gluon GPD contributions and their total.
The result are similar to the cross sections 
in Fig.\,\ref{fig:pi+-cross-sections}
for the reaction $\nu  p \to \ell^-  p' \, \pi^+$.
The cross section is dominated by the gluon terms;
however, there is a tendency that the fraction of 
the quark-GPD contribution increases at large $Q^2$
as noted in Fig.\,\ref{fig:pi+-cross-sections}.

At the Fermilab, $\nu_\mu$ and $\bar \nu_\mu$ scattering experiments
are possible with the hydrogen (H) and nuclear (A) targets by using
the high-intensity LBNF beam \cite{lbnf-beam,kp-2021}. 
Using the proton target H, we could investigate
the pion-production processes
\begin{align}
p:\ \, 
\nu_\mu \, p \to \mu^- p \, \pi^+ , \ \ 
\bar \nu_\mu \, p \to \mu^+ p \, \pi^- , \ \ 
\bar \nu_\mu \, p \to \mu^+ n \, \pi^0 .
\nonumber
\end{align}
If a nuclear target A, such as carbon and argon, is used, 
it is possible to access the additional reactions
with the neutron:
\begin{align}
n:\ \, 
\nu_\mu \, n \to \mu^- n \, \pi^+, \ \ 
\nu_\mu \, n \to \mu^- p \, \pi^0, \ \ 
\bar \nu_\mu \, n \to \mu^+ n \, \pi^- ,
\nonumber
\end{align}
where nuclear corrections should be applied properly.
Because of the isospin symmetry, we expect that
the neutron cross sections are related to the proton 
cross sections theoretically as 
\begin{alignat}{2}
\text{Fig.\,\ref{fig:pi+-cross-sections}}: & \ \  & 
\sigma(\bar \nu_\mu \, n \to \mu^+ n \, \pi^-) 
  & =\sigma (\nu_\mu \, p \to \mu^- p \, \pi^+) ,
\nonumber \\
\text{Fig.\,\ref{fig:pi+-cross-sections-n}}: & \ \  &
\sigma(\nu_\mu \, n \to \mu^- n \, \pi^+) 
  & =\sigma (\bar \nu_\mu \, p \to \mu^+ p \, \pi^-) ,
\nonumber \\
\text{Fig.\,\ref{fig:pi0-cross-sections}}: & \ \ &
\sigma (\nu_\mu \, n \to \mu^- p \, \pi^0)
  & =\sigma (\bar \nu_\mu \, p \to \mu^+ n \,\pi^0) ,
\end{alignat}
where the corresponding numerical results are shown
in Figs.\,\ref{fig:pi+-cross-sections}, 
\ref{fig:pi+-cross-sections-n}, and 
\ref{fig:pi0-cross-sections}.
The third cross section $\sigma (\nu_\mu \, n \to \mu^- p \, \pi^0)$
is discussed in the next subsection.
Using these cross sections, we can estimate all the reaction
cross sections numerically for the pion ($\pi^\pm$, $\pi^0$) productions
in the neutrino and antineutrino reactions at Fermilab
by taking into account nuclear modifications 
if the target is a nucleus.
We also mention that the transverse form factor component 
of the GPD $\tilde H^q$ is constrained by the axial form factor 
measured in neutrino reactions, for example, 
by the recent MINER$\nu$A experiment \cite{minerva-2023}.

\subsection{Cross sections for the $\pi^0$ production}
\label{pi0}

Next, we discuss the $\pi^0$-production cross section 
in neutrino scattering. Here, the cross section is sensitive
to the quark GPD combination
$[  \{ F^u (x,\xi,t) - F^d (x,\xi,t) \}
  + \{ F^u (-x,\xi,t) - F^d (-x,\xi,t) \} ]/\sqrt{2}$
as shown in Eq.\,(\ref{eqn:pi0-Fq}).
In the charged-current neutrino scattering,
there is no leading-gluon term, as shown in Fig.\,\ref{fig:neutrino-pi-gpd},
for the $\pi^0$ production because of the charge conservation
$W^\pm \not\to \pi^0$. Therefore, it is suitable for determining
the quark GPDs. Furthermore, the pion- and rho-pole GPDs are
symmetric under the transformation $x \to -x$ as obvious
in Eqs.\,(\ref{eqn:pion-pole}) and (\ref{eqn:rho-pole}),
there are pion and rho-pole GPD contributions
($[  \{ F^u (x,\xi,t) - F^d (x,\xi,t) \}
  + \{ F^u (-x,\xi,t) - F^d (-x,\xi,t) \} ]_{\pi,\rho\text{-poles}}/\sqrt{2} $)
to the cross section. It means that the $\pi^0$-production 
could be suitable for determining the quark GPDs including
the pion- and rho-pole terms.

\begin{figure}[b]
 \vspace{-0.40cm}
\begin{center}
   \includegraphics[width=8.5cm]{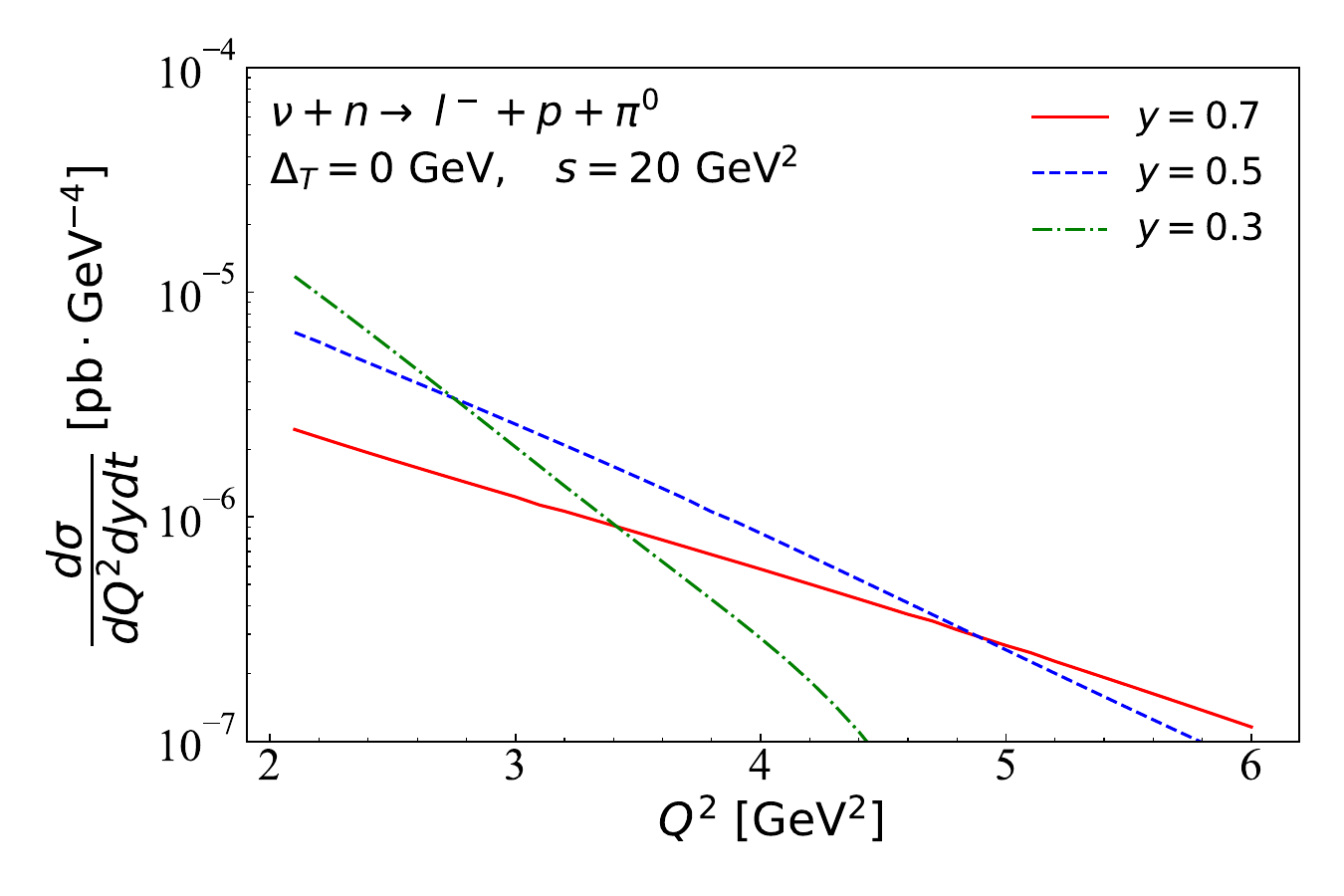}
\end{center}
\vspace{-0.9cm}
\caption{$\pi^0$ production cross sections
for $\nu n \to \ell^- p \, \pi^0$
at $\Delta_T =0$ GeV and $s=20$ GeV$^2$. }
\label{fig:pi0-cross-sections}
%
\vspace{-0.20cm}
\begin{center}
   \includegraphics[width=8.5cm]{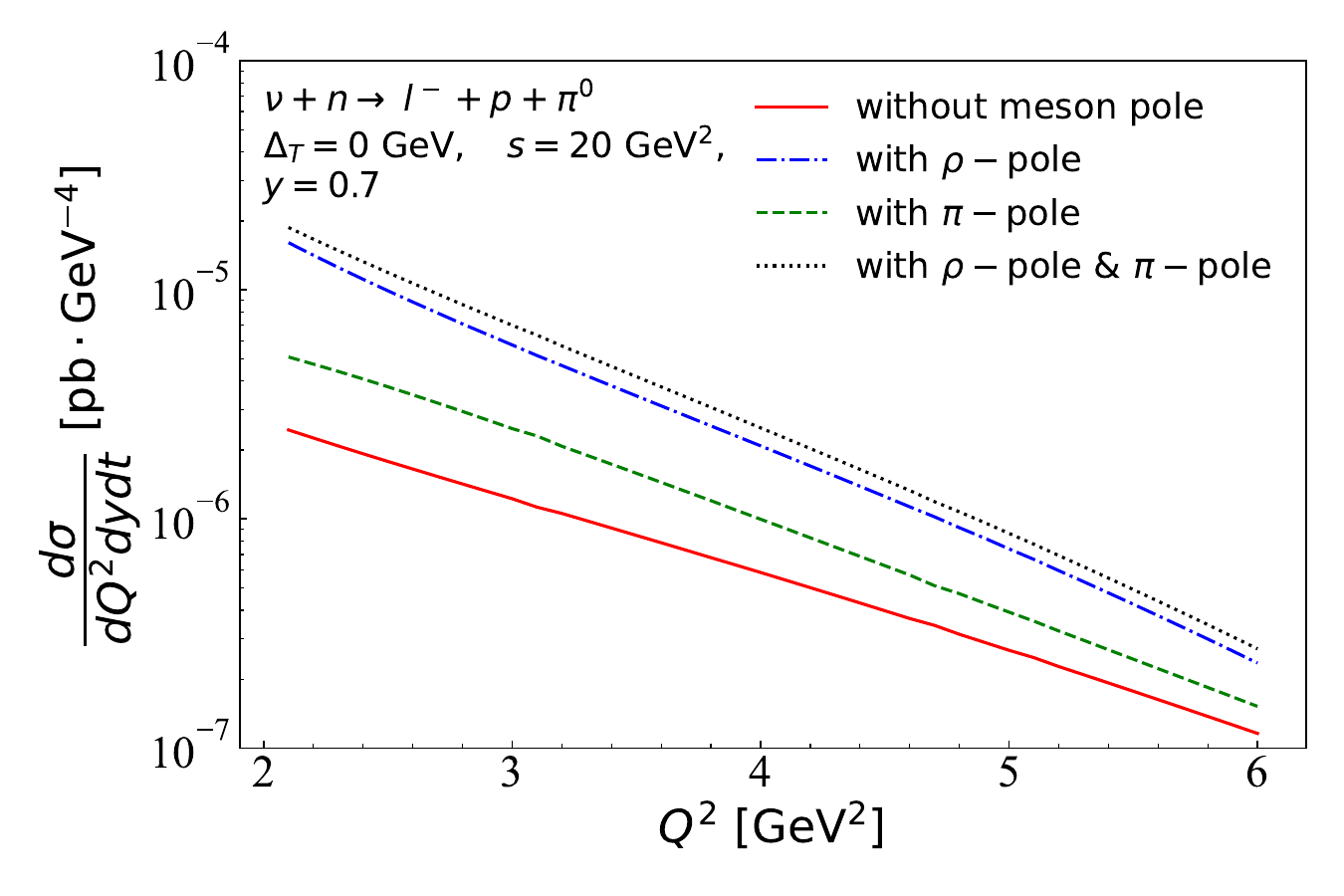}
\end{center}
\vspace{-0.9cm}
\caption{Pion- and rho-pole effects on the $\pi^0$ production 
cross section for $\nu n \to \ell^- p \, \pi^0$
at $y=0.7$, $\Delta_T =0$ GeV and $s=20$ GeV$^2$. }
\label{fig:pi0-cross-sections-poles}
\vspace{-0.00cm}
\end{figure}

\begin{figure*}[t]
 \vspace{-0.00cm}
\begin{center}
   \includegraphics[width=17.0cm]{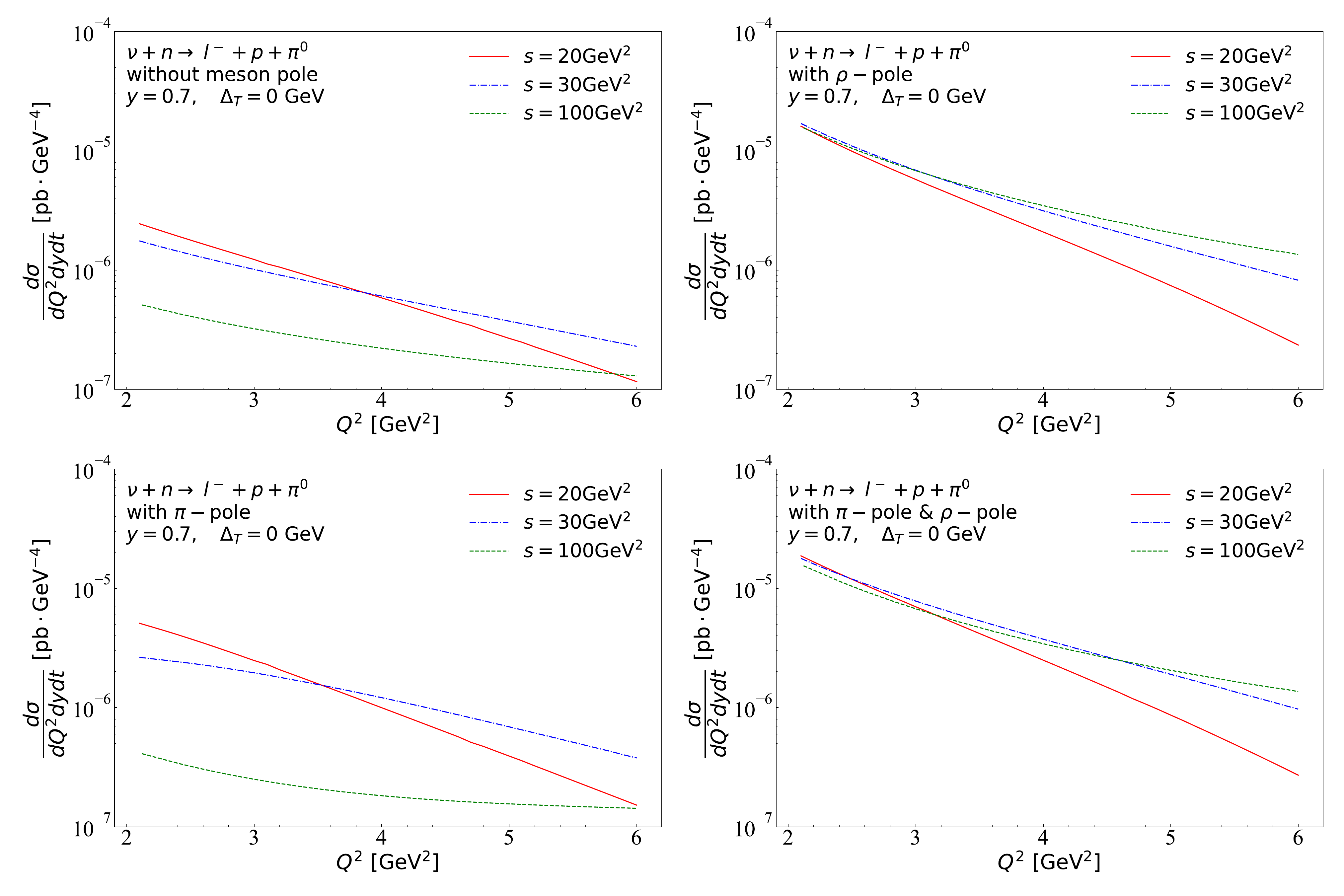}
\end{center}
\vspace{-0.7cm}
\caption{
Center-of-mass-energy squared dependence of 
the $\pi^0$-production cross section
for $\nu n \to \ell^- p \, \pi^0$
by taking $s=20$, 30, and 100 GeV$^2$
with or without the pion- and rho-pole GPDs.
}
\label{fig:pi0-cross-energy-dep}
\vspace{-0.30cm}
\end{figure*}

Calculated $\pi^0$ cross sections 
for the process  $\nu + n \to \ell^- + p + \pi^0$
are shown in Fig.\,\ref{fig:pi0-cross-sections}
by taking the kinematical conditions, 
$\Delta_T =0$ GeV and $s=20$ GeV$^2$.
Three curves are shown for $y=0.3$, 0.5, and 0.7.
The largest contribution comes from the GPDs $H^q$,
and subsequent one is $\tilde H^q$, and then
$E^q$ and $\tilde E^q$ in order
as the same way with the $\pi^+$ production.
Here, the pion- and rho-pole terms are not included.

Next, including the pion- and rho-pole GPD contributions,
we show the $\pi^0$ cross sections
for the process  $\nu + n \to \ell^- + p + \pi^0$
by taking $y=0.7$, $\Delta_T =0$ GeV and $s=20$ GeV$^2$
in Fig.\,\ref{fig:pi0-cross-sections-poles}.
Four cross sections are shown. 
The solid curve indicates the cross section without the pion 
and rho-pole GPDs, 
the dashed curve does the one by including only the pion-pole GPDs,
the dashed-dot curve does by including only the rho-pole GPDs, and
the dotted curve does by including the the pion and rho-pole GPDs.
It is obvious that the pion- and rho-pole effects increase 
the $\pi$-production cross section significantly.
The increase is much larger by adding the rho pole than the pion pole.
It is simply because the rho-pole GPDs exist in the function $H$,
which is the largest component in the pion-production cross sections.
In this way, we found that the pion- and rho-pole GPDs could be
studied in the $\pi^0$ production process. 
However, such pronounced meson-pole GPD effects should be 
re-estimated by a new global analysis including the data
of charged-lepton scattering as we mentioned in Sec.\,\ref{pi+}.
The GPDs in the ERBL region, namely the meson-pole terms, 
have never been investigated experimentally, but they could be studied
by the neutrino reactions, especially by the $\pi^0$ production processes.
These reactions are totally different from the hadronic reaction 
under consideration at J-PARC for probing the ERBL region
\cite{kss-2009,J-PARC-GPDs,transition-GPDs}.

The energy dependence of the $\nu n \to \ell^- p \, \pi^0$
cross section is shown in Fig.\,\ref{fig:pi0-cross-energy-dep}
by taking $s=20$, 30, and 100 GeV$^2$ with or without
the pion- and rho-pole GPDs.
Without the meson-pole GPDs, the cross section decreases
with increasing $s$ as shown in the upper-left figure
as a general tendency, although the details depends also
on $Q^2$ as the $s=30$ GeV$^2$ cross section becomes larger
than the $s=20$ GeV$^2$ one at large $Q^2$.
This general tendency is opposite to the increasing cross section 
of $\nu p \to \ell^- p' \, \pi^+$ with $s$
in Fig.\,\ref{fig:pi+crross-energy-dep}.
It is because the quark GPDs do not increase
as rapidly as the gluon GPDs as $x$ decreases with
the increasing $s$. 
This energy dependence stays the same even if the pion-pole GPDs
are added as shown in the lower-left figure,
although the overall cross sections 
become larger at $s=20$ and 30 GeV$^2$.
If the rho-pole GPDs are added as shown in the upper- 
and lower-right figures, the cross sections become much larger
and the energy dependence is much different.
As $s$ increases to $s=30$ and 100 GeV$^2$, 
the cross section becomes much larger
than the one of $s=20$ GeV$^2$, 
especially in the large $Q^2$ region.
The difference between the pion- and rho-pole GPD effects
comes from the fact that the rho-pole GPDs exist 
in the main component $H^q$ (and subleading one $E^q$)
and the pion-pole GPDs are in the subleading $\tilde E^q$
in the cross section.
These results indicate again that the neutrino reactions
could be used for probing the ERBL region of the GPDs.

As seen in the cross-section figures, the cross sections are shown
in the $Q^2$ range from 2 GeV$^2$ to 6 GeV$^2$.
In the low $Q^2$ region at $Q^2 \sim 2$ GeV$^2$, there could be significant 
corrections due to higher-twist and higher-order-$\alpha_s$ effects.
For example, it is interesting to include the these effects
in Eq.\,(\ref{eqn:quark-gluon-GPD-integral}) in the sense that 
the cross section becomes sensitive to
the functional form of the pion distribution amplitude $\Phi_\pi (x)$
as shown in Ref.\,\cite{KSS-2014}.
In the cross sections integrated over $\phi$, the higher-twist terms
$\sigma_{TT}$ and $\sigma_{LT}$ does not contribute. 
In the unintegrated pion-production cross sections in the charged-lepton 
scattering, the higher-twist effects were calculated and they are 
sizable magnitude at small $Q^2(=$1--2 GeV$^2$) in comparison 
with the leading-twist effects 
\cite{GHLM-2010,pion-production-electron,Goloskokov:2022mdn}.
In fact, an analysis result for the pion electroproduction cross sections
indicates that the higher-twist contributions could be 
comparable to the leading one in magnitude 
at $Q^2=$2 GeV$^2$ and $s=20$ GeV$^2$.
The cross sections $\sigma_{TT}$ and $\sigma_{LT}$ are expressed
by the transversity GPDs
\cite{GHLM-2010,pion-production-electron,Goloskokov:2022mdn}; 
however, there is no study on the meson-pole effects
on the transversity GPDs as far as we are aware.
Because the $\sigma_{TT}$ and $\sigma_{LT}$ are sizable in magnitude
at small $Q^2$, it could be important to include the meson-pole effects
also on the transversity GPDs in future analysis.

\begin{figure}[h]
 \vspace{-0.30cm}
\begin{center}
   \includegraphics[width=7.5cm]{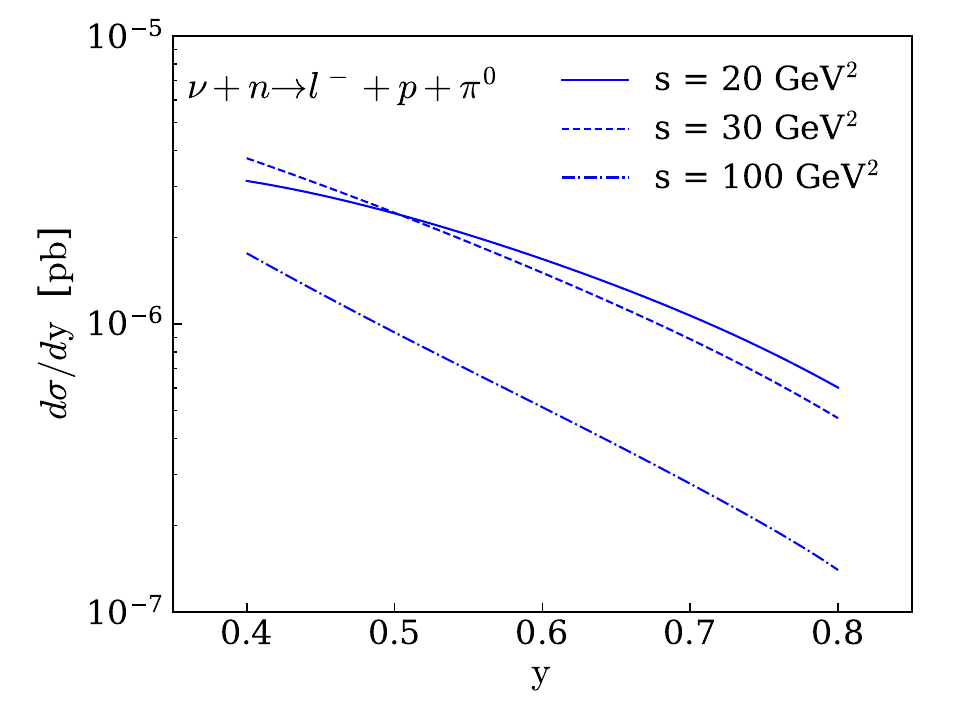}
\end{center}
\vspace{-0.7cm}
\caption{Single differential cross section $d\sigma /dy$
is shown for $\nu n \to \ell^- p \, \pi^0$
by taking $s=20$, 30, and 100 GeV$^2$.}
\label{fig:dsdy-pi0}
\vspace{-0.00cm}
\end{figure}

The single differential cross section $d\sigma /dy$
is shown for the process  $\nu n \to \ell^- p \, \pi^0$
by integrating the cross section over $Q^2$ and $t$
at $s=20$, 30, and 100 GeV$^2$ in Fig.\,\ref{fig:dsdy-pi0}.
The cross section is about 
$1\times 10^{-6}$--$4\times  10^{-6}$\,pb
at $y=0.4$--$0.5$.

In general, neutrino DIS measurements have been valuable 
in determining light-quark flavor-dependent distributions,
the valence-quark distributions, 
and the strange-quark distribution. 
The situation could be same for the GPD determination,
and future neutrino reaction experiments will provide
valuable information on the gluon GPDs and also the quark GPDs, 
especially on the flavor dependence.
One may note that it is not easy to achieve precise
measurements in neutrino reactions because the neutrino beam 
has a wide energy distribution instead of the monochromatic 
one in the charged-lepton beam. 
However, a detector with a wide angular acceptance and 
a low-momentum threshold could make it possible to investigate
the GPDs in the neutrino reactions \cite{lu-2024}.
Although there is no experiment for the GPDs in neutrino scattering,
such a project is under consideration at Fermilab \cite{kp-2021}
by using the high-energy neutrino beam of about 10 GeV \cite{lbnf-beam},
and it may be possible also at the nuSTORM \cite{nustorm,lu-2024}.
We hope that these experimental projects will be realized and 
that it will lead to a better understanding on the GPDs.

\section{Summary}
\label{summary}

We investigated the pion-production processes 
in neutrino reactions for determining the nucleon GPDs.
Using the GK parametrization for the GPDs, we showed
the effects of the gluon and quark GPDs in the process
$\nu + p \to \ell^- + p' + \pi^+$.
The largest contribution to the $\pi^+$ production comes
from the GPD $H^g$, then subsequent contributions are
$H^q$, $\tilde H^q$, and $E^q$, $\tilde E^q$, $E^g$ in order
by showing each term in the cross section.
The pion-pole GPDs are large in the function $\tilde E$,
and the rho-pole GPDs exist as large components in 
the functions $H$ and $E$. We showed these-pole 
effects in the $\pi^+$-production cross sections.
We found that the GPD $E^g$ effects are small in the $\pi^+$-production
cross section according to the current parametrization for $E^g$.
In the $\pi^0$ production $\nu + n \to \ell^- + p + \pi^0$, 
the quark GPDs can be investigated. In particular, the pion- and rho-pole
contributions are large in the $\pi^0$ cross section,
which indicates that this process could be used
for finding the GPDs in the ERBL region.
However, the additional rho-pole GPDs should be tested
by a new global analysis with the charged-lepton data.
We also showed the dependence of the cross sections
on the center-of-mass energy squared
by calculating the cross sections at $s=20$, 30, and 100 GeV$^2$.
As $s$ increases, the $\nu + p \to \ell^- + p' + \pi^+$
cross section increases, whereas 
the $\nu + n \to \ell^- + p + \pi^0$ one 
decreases or increases with $s$, depending mainly 
on the meson-pole GPDs and also on $Q^2$.
This difference between the two processes comes from 
the difference between the gluon and quark GPDs as $x$ decreases. 
In general, the gluon GPDs increase more rapidly
than the quark GPDs as $x$ becomes smaller.
Future neutrino GPD projects will be valuable
also for finding the flavor dependence on the quark GPDs
in the same manner with the unpolarized PDF determination.

We also note that there were related works recently 
on the electroweak process $\ell N \to \nu \pi N'$ 
to investigate the GPDs at the future EICs \cite{electroweak-GPDs}.
Our current studies could be extended to such processes.

\begin{acknowledgements}
The work was partially supported by Strategic Priority Research Program 
of the NSFC under Grant No. 12293060 and No.12293061, 
 Chinese Academy of Sciences under Grant No. XDB34030301,
and the CAS president's international fellowship initiative under
Grant No. 2022VMA0003.
It was also partially supported by the Japan Society for the Promotion 
of Science (JSPS) Grants-in-Aid for Scientific Research (KAKENHI) 
Grant Number 19K03830.
The authors would like to thank B. Pire, L. Szymanowski, and J. Wagner
for suggestions on pion-production cross sections,
R. Petti for communications on the Fermilab neutrino GPD experiment,
and Wen-Chen Chang, S. V. Goloskokov, and P. Kroll
for suggestions on the GPD parametrization.
\end{acknowledgements}


\end{document}